\documentclass[utf8,10pt]{frontiersinFPHY_FAMS} 
\usepackage{url,hyperref,lineno,microtype,subcaption}
\usepackage[onehalfspacing]{setspace}
\usepackage{graphicx}
\usepackage{dcolumn}
\usepackage{bm}
\usepackage[utf8]{inputenc}
\usepackage[T1]{fontenc}
\usepackage{mathptmx}
\usepackage{verbatim}
\usepackage{fdsymbol}
\usepackage{graphicx}
\usepackage{sidecap}
\usepackage{placeins}
\usepackage{xcolor}
\usepackage{amsfonts,amsthm,amsmath,amssymb,amscd,graphicx}
\usepackage{mathtools}
\usepackage{hyperref}
\usepackage{amsmath}
\usepackage{amssymb}
\usepackage{bbold}
\usepackage{float}
\usepackage{hyperref}
\usepackage[onehalfspacing]{setspace}

\def\keyFont{\fontsize{8}{11}\helveticabold }
\def\firstAuthorLast{Pusuluri {et~al.}} 
\def\Authors{Krishna Pusuluri\,$^{1,2,*}$, Huiwen Ju\,$^{1}$ and Andrey L Shilnikov\,$^{1,3}$}
\def\Address{$^{1}$ Neuroscience Institute, Georgia State University, Atlanta, Georgia, USA\\
$^{2}$Tri-Institutional Center for Translational Research in Neuroimaging and Data Science (TReNDS), Georgia State University, Georgia Institute of Technology, and Emory University, Atlanta, Georgia, USA\\
$^{3}$Department of Mathematics \& Statistics, Georgia State University, Atlanta, Georgia, USA}
\def\corrAuthor{Krishna Pusuluri, 55 Park Place NE, Atlanta, GA 30303, USA}
\def\corrEmail{kpusuluri1@gsu.edu}

\begin{document}
\onecolumn
\firstpage{1}

\title[Emergent network bursting in a 4-cell CPG]{Hierarchical emergence of network bursting in a four-cell central pattern generator model}


\author[\firstAuthorLast ]{\Authors} 
\address{} 
\correspondance{} 

\extraAuth{}%

\maketitle
\begin{abstract}

\tiny
 \keyFont{ \section{Keywords:} central pattern generators, hierarchical emergence, half-center oscillators, neural network dynamics, symbolic encoding, \textit{Dendronotus iris}} 
\end{abstract}

How can a neural circuit rhythmically burst when none of its constituent neurons can endogenously do so? We address this question through a bottom-up reconstruction of a 4-cell neural circuit modeled after the swim central pattern generator (CPG) of the sea slug \textit{Dendronotus iris}. We first map the intrinsic regimes of a swim interneuron (SiN) model neuron and show that slow mutual inhibition can generate anti-phase bursting in a half-center oscillator (HCO) assembled from tonic-spiking or quiescent cells. A slow--fast phase-space deconstruction explains this pairwise rhythm as a release mechanism. We then move on to CPG parametrization, in which cells 1 and 2 are quiescent, while cells 3 and 4 are tonic spikers but their isolated HCO can only exhibit tonic spiking or suppression. Bursting therefore does not arise at either the cellular or HCO level. It appears only after the two modules are assembled into the complete 4-cell network, where cross excitation, cross inhibition, and rectified electrical coupling act together to support a robust network rhythm-generation. Event-based symbolic encoding and GPU-parallel parameter sweeps show that this higher-order collective state occupies extended parameter domains rather than a single tuned point, and they quantify how the network interactions reshape those domains and the spike counts per burst. Symbolic sweeps identify the healthy range of activity regimes (periodic sequences) and transition boundaries; Lempel--Ziv complexity is used as a descriptor of aperiodic symbolic output rather than as proof of chaos or dynamical instability. Our results establish a hierarchy of rhythm generation---from intrinsic cell dynamics, through conditional HCO bursting, to bursting that emerges only in the fully assembled 4-cell CPG---and provide experimentally accessible predictions for perturbing its chemical and electrical couplings.

\section{Introduction}

Rhythmic motor output is often attributed either to intrinsic bursting in individual neurons or to oscillations generated by small recurrent circuit motifs composed of neurons coupled by chemical and electrical synapses. Central pattern generators (CPGs), however, are organized hierarchically: neurons form half-center oscillators (HCOs), and HCO-like modules are coupled into larger networks. This organization raises a more specific question than whether coupling can modify an existing rhythm: can rhythmic bursting emerge only at a higher level of circuit assembly, when neither the component neurons nor the relevant isolated modules can burst?

CPGs generate the rhythmic neural activity underlying walking, crawling, swimming, respiration, cardiac activity, and digestion without requiring rhythmic sensory feedback or descending input \citep{bal1988pyloric,bulloch1992reconstruction,calin2007parameter,frost1996single,katz2016evolution,kristan2005neuronal,marder1994invertebrate,marder1996principles,miller1985neural,milo2002network,newcomb2012homology,prinz2019rhythmic,rabinovich2006dynamical,sherwood2011synaptic,sporns2004motifs}. A common CPG motif is the HCO, in which two neurons inhibit one another and alternate their activity. Multiple such motifs interact through chemical and electrical synapses in the swim CPGs of the sea slugs \textit{Melibe leonina} and \textit{Dendronotus iris} \citep{newcomb2012homology,alacam2015making,brown1911intrinsic,jalil2013toward,katz1998comparison,katz2017neural,sakurai2014two,sakurai2011distinct,sakurai2016central,sakurai2017artificial,sakurai2011different}. The dynamical contribution of each hierarchical level cannot be inferred from connectivity alone. Isolated neurons may be quiescent or tonic spikers, mutual inhibition may or may not generate alternating bursts, and interactions between modules may produce a collective rhythm that is absent from every isolated subsystem.

Here, we distinguish three levels of rhythm generation. \emph{Intrinsic bursting} occurs in an isolated neuron. \emph{HCO-emergent bursting} is generated through mutual inhibition between neurons that do not burst intrinsically. \emph{Higher-order network-emergent bursting} appears only when nonbursting lower-order components interact within the complete circuit. The SiN model allows us to establish the transition from intrinsic cellular activity to HCO-emergent bursting and to examine the underlying slow--fast release mechanism. The \textit{Dendronotus} parameterization then provides a critical comparison: cells~3 and~4 are intrinsic tonic spikers, yet their isolated HCO produces only tonic spiking or suppression, whereas cells~1 and~2 are quiescent. Nevertheless, the assembled four-cell circuit generates bursting through the combined action of cross excitation, cross inhibition, and rectifying electrical coupling.

Establishing this hierarchy requires more than a single hand-selected simulation. Conductance-based circuits occupy large parameter spaces, and qualitatively similar outputs can arise from substantially different combinations of intrinsic and synaptic parameters \citep{prinz2004similar}. Optimization can identify parameter sets that reproduce a target trajectory, but its results depend on the selected objective function, optimization algorithm, and initial conditions \citep{bourahmah2024error,friedrich2014flexible}. Numerical continuation and Lyapunov analysis provide more detailed local dynamical information, but applying these methods exhaustively across large network parameter spaces can be computationally prohibitive \citep{barrio2014,barrio2011parameter,pusuluri2019symbolic}. We therefore use event-based symbolic coarse graining of voltage trajectories, implemented within the Deterministic Chaos Prospector framework, to classify network activity across GPU-parallel parameter sweeps \citep{pusuluri2019symbolic,pusulurihomoclinicCNSNS,pusuluri2017unraveling,pusuluri2018homoclinic}. This computational approach identifies periodic symbolic patterns, aperiodic or complex output, suppression, tonic spiking, and bursting while retaining experimentally relevant observables such as the number of spikes per burst.

Our analysis follows the hierarchy of circuit assembly. We first map the activity regimes of isolated model neurons and identify where slow mutual inhibition recruits nonbursting cells into an alternating HCO rhythm. We then use a slow--fast phase-space representation to characterize the release mechanism underlying this conditional pairwise bursting. Next, we show that the corresponding isolated HCO under the \textit{Dendronotus} parameterization does not burst. Finally, we demonstrate that the fully assembled circuit recovers bursting across extended parameter regions and determine how excitation, inhibition, and electrical coupling reshape those regions. The resulting symbolic maps are used to identify activity regimes and transition boundaries; they are not interpreted, by themselves, as evidence of formal bifurcations, chaos, or topological equivalence between attractors.

\section{Methods}
\subsection{Symbolic representation of neuron and network dynamics}
\begin{figure}[t!]
\centering
\includegraphics[width=.7\textwidth]{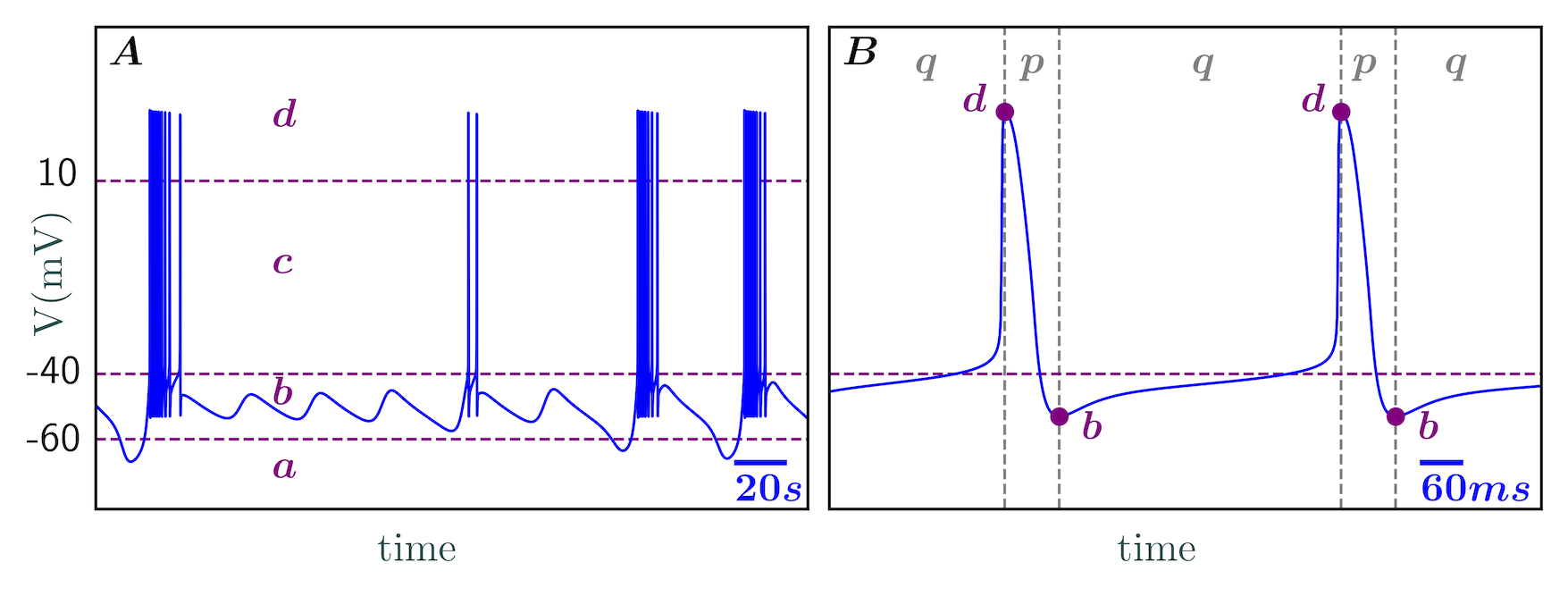}
\caption{Symbolic encoding of a chaotic bursting trajectory with subthreshold oscillations in the
SiN model. (A) Voltage-based partition of the trajectory using the thresholds
$V_{\rm bins}=[-60,-40,10]~\mathrm{mV}$, indicated by the purple dashed
lines. These three thresholds divide the voltage range into four symbolic
intervals:
$a$ for $V\leq-60~\mathrm{mV}$,
$b$ for $-60<V\leq-40~\mathrm{mV}$,
$c$ for $-40<V\leq10~\mathrm{mV}$, and
$d$ for $V>10~\mathrm{mV}$.
(B) Magnification of a short intraburst segment containing two spikes.
The local voltage maxima and minima are marked by purple dots. Under the
voltage-based partition, these successive extrema generate the symbolic
sequence $(dbdb)$, shown in purple. A complementary temporal partition is constructed from the time intervals
between consecutive events using the threshold
$T_{\rm bins}=[100]~\mathrm{ms}$. Intervals of duration
$\Delta t\leq100~\mathrm{ms}$ are assigned the symbol $p$, whereas intervals
with $\Delta t>100~\mathrm{ms}$ are assigned the symbol $q$. The temporal
encoding of the displayed segment is $(qpqpq)$, shown in gray; the gray
dashed lines delimit the corresponding event intervals. Interleaving the
temporal and voltage symbols produces the combined sequence
$(qdpbqdpbq)$, which retains information about both the event amplitudes
and their timing}\label{fig:plantTrajectorySymbols}
\end{figure}

Our symbolic encoding is an event-based coarse-graining procedure motivated
by experimental neurophysiology, in which membrane-voltage recordings are
often the principal observable. A simple binary partition can record the
presence or absence of a spike within consecutive short time bins. Here, we
use a richer event-based partition that detects local voltage maxima and
minima and assigns symbols according to both their amplitudes and the time
intervals between successive events. This construction allows us to
distinguish quiescence, tonic spiking, spike addition, different bursting
waveforms, and mixed-mode activity. The resulting symbolic word provides a
compact description of the voltage trajectory under a specified partition;
it is neither unique nor assumed to constitute a generating partition of the
full conductance-based state space. Accordingly, interpretations of the
symbolic sequences are calibrated against representative voltage traces and,
when available, independent phase-space or bifurcation information.

Figure~\ref{fig:plantTrajectorySymbols} illustrates an irregular
mixed-mode bursting trajectory generated by the original SiN model
\citep{alacam2015making}, which is described in the following sections.
The trajectory exhibits varying numbers of spikes per burst and
subthreshold oscillations between successive bursts. We detect the local
voltage maxima and minima, indicated by the colored dots in
Fig.~\ref{fig:plantTrajectorySymbols}, and calculate the time intervals
between successive detected events, delimited by the vertical gray dashed
lines. Voltage and temporal partitions, denoted by
$V_{\mathrm{bins}}$ and $T_{\mathrm{bins}}$, respectively, are then used to
map the event amplitudes and inter-event intervals onto symbolic sequences.

For the voltage thresholds $V_{\mathrm{bins}}=[-60,-40,10]mV$, the voltage range is divided into four symbolic intervals: 
(a): $V\leq -60mV$; (b): $-60<V\leq -40mV$; (c): $-40<V\leq 10mV$, and (d): $V>10mV$.
These symbols correspond approximately to quiescence or burst termination,
subthreshold oscillations, the burst plateau, and action-potential peaks,
respectively. The physiological interpretation of each symbol depends on
whether the detected event is a local maximum or minimum and on its position
within the surrounding symbolic sequence.

For the temporal threshold $T_{bins}=[100]ms$,
an inter-event interval $\Delta t\leq 100ms$ is assigned the symbol
$p$, whereas an interval $\Delta t>100ms$ is assigned the symbol
$q$. Thus, $p$ and $q$ represent relatively short and long intervals,
respectively, between successive voltage extrema. For the magnified two-spike segment shown in
Fig.~\ref{fig:plantTrajectorySymbols}B, the voltage partition produces the
sequence $(dbdb)$, while the temporal partition produces $(qpqpq)$. The
latter includes the intervals immediately preceding and following the four
displayed voltage events. Interleaving the temporal and voltage symbols
yields the combined sequence $(qdpbqdpbq)$, which preserves information
about both event amplitudes and timing. Using an overbar to denote periodic repetition, a tonic-spiking trajectory with the same event structure can be represented by
$\overline{db}$ under the voltage partition,
$\overline{qp}$ under the temporal partition, or
$\overline{qdpb}$ under the combined partition. Similarly, a regular
three-spike burst without intervening subthreshold oscillations can be
represented by the voltage sequence $\overline{dbdbda}$, where the terminal
symbol $a$ denotes the hyperpolarized minimum associated with burst
termination. Bursts containing subthreshold oscillations or other
mixed-mode features generate additional symbols, particularly $b$ and $c$,
whose ordering distinguishes these more complex waveforms. 

\begin{figure*}[t!]
\centering
  {\includegraphics[width=0.5\textwidth]{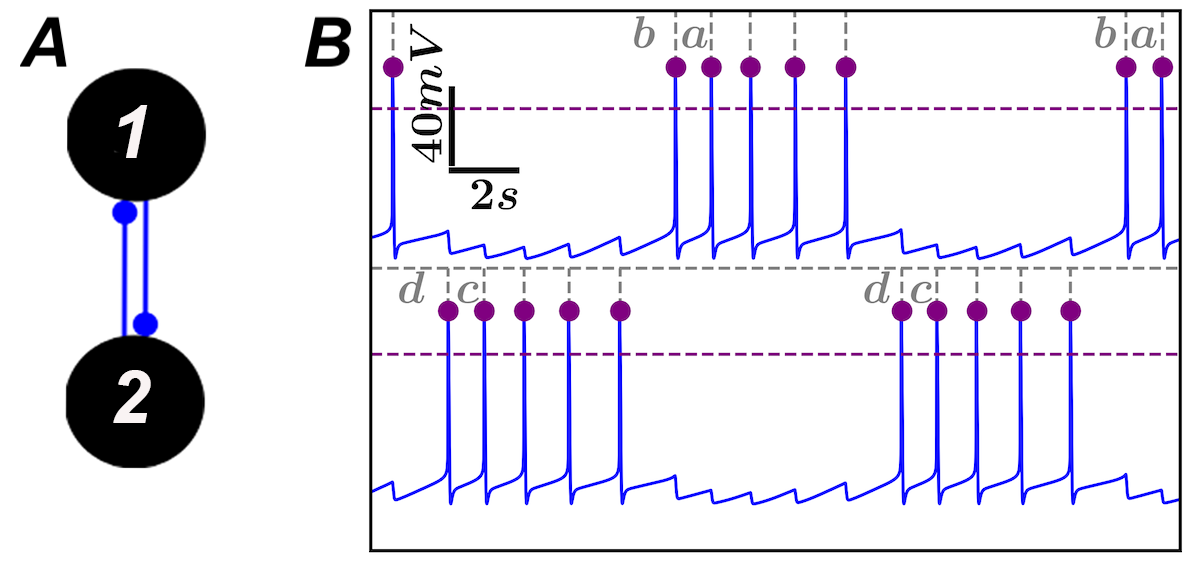}}  
\caption{Symbolic encoding of anti-phase bursting in a half-center oscillator (HCO)
composed of two mutually inhibitory SiN model cells, as illustrated in
panel~(A). A voltage threshold of
$V_{bins}=10mV$ is used to detect spike maxima above the
threshold, marked by purple dots. The intervals $\Delta t$ between successive
detected events, delimited by the gray vertical lines, are classified using
the temporal threshold $T_{bins}=2.5s$. For cell~1, intervals with $\Delta t\leq 2.5~s$ are assigned the
symbol $a$, whereas intervals with $\Delta t>2.5~\mathrm{s}$ are assigned
the symbol $b$. For cell~2, the corresponding short and long intervals are
assigned the symbols $c$ and $d$, respectively. Thus, the first detected
spike of a burst in cell~1 follows a long interburst interval and is assigned
the symbol $b$, while the second spike follows a short intraburst interval
and is assigned the symbol $a$. The analogous events in cell~2 are assigned
the symbols $d$ and $c$, respectively.
Only the first two spikes of each burst are retained in the symbolic
description. Additional intraburst spikes, which would otherwise receive the
short-interval symbols $a$ or $c$, are omitted to prevent ambiguities when
burst durations overlap between the two cells. With this convention, the
repeating anti-phase bursting rhythm of the HCO is represented by the
periodic symbolic sequence $\overline{dcba}$.}\label{fig:PlantHCOSymbolsNetwork}
\end{figure*}

The same event construction can incorporate cell-specific symbol sets and thereby describe a network rhythm. Fig.~\ref{fig:PlantHCOSymbolsNetwork} illustrates the encoding of an anti-phase HCO. We use $V_{bins}=[10]mV$ to detect spike maxima and $T_{bins}=[2.5]s$ to distinguish the first two spikes of each burst in cells $1$ and $2$. Later spikes within a burst are omitted from the word so that modest overlap of burst durations does not create spurious phase differences. The repeating anti-phase rhythm is therefore encoded as $(\overline{dcba})$. We also record the mean spikes per burst when a rhythm is detected and total spike counts otherwise. These criteria are tailored to the dedicated alternating rhythms studied here; a multifunctional circuit would require partitions that explicitly discriminate its competing phase-locked states. For the \textit{Dendronotus} model we use $V_{bins}=[10]mV$ for spike maxima and $T_{bins}=[1]s$ for burst detection.

\subsection{Biparametric sweeps of neuron and network dynamics}
\begin{figure}[ht!]
\centering
\includegraphics[width=0.8\textwidth]{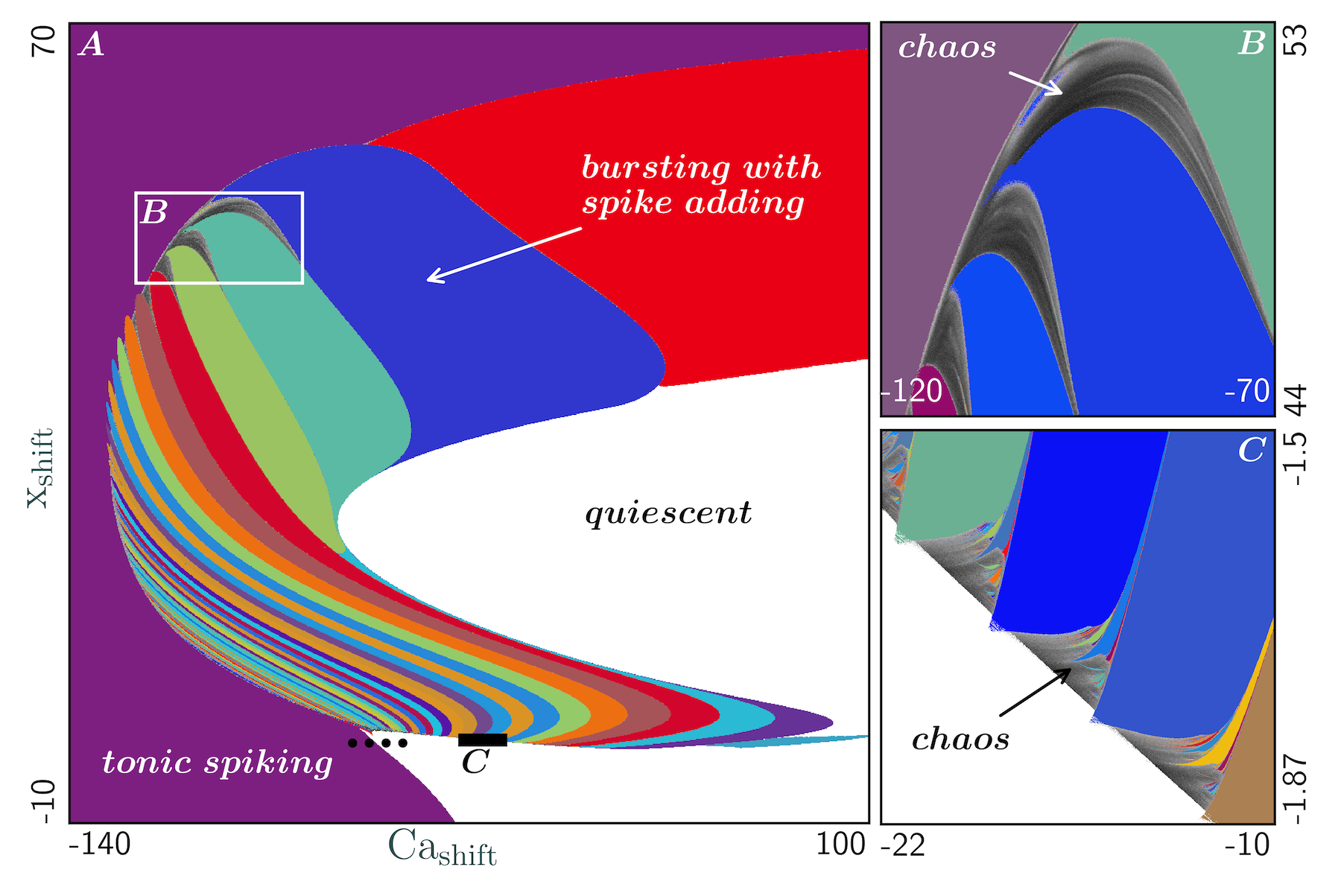}
\caption{(A) Two-parameter sweep of the SiN model showing the activity regimes and
transitions produced by varying the intrinsic cellular parameters.
Representative voltage trajectories displaying bursting, quiescence, and
tonic spiking are presented in
Figs.~\ref{fig:PlantAllTrajectories}A and~B.
Chaotic spiking and bursting occur within the gray regions near the boundary
between the regular bursting and tonic-spiking regimes. This region is
magnified in panel~(B), and a representative trajectory is shown in
Fig.~\ref{fig:PlantAllTrajectories}D.
Chaotic mixed-mode oscillations occur near the boundary between the bursting
and quiescent regimes. This region is magnified in panel~(C), with a
representative trajectory shown in
Fig.~\ref{fig:PlantAllTrajectories}C.
The four black dots located near the lower boundary between the tonic-spiking
and quiescent regimes in panel~(A) mark the parameter sets used to study
emergent bursting in an HCO composed of cells that do not burst
intrinsically. From left to right, these points correspond to panels~A--D
of Fig.~\ref{fig:PlantHCOGrid}.}\label{fig:PlantSweep}
\end{figure}

Two-parameter sweeps such as that shown in Fig.~\ref{fig:PlantSweep} are generated by integrating voltage trajectories over a parameter grid using a fixed-step fourth-order Runge--Kutta method. Trajectories are evaluated in parallel on CUDA GPU threads. After the transient portion of each trajectory is discarded, the remaining output is encoded using the specified $V_{bins}$ and $T_{bins}$ partitions and the corresponding network-level classification criteria.

Periodic symbolic words are normalized with respect to cyclic permutation before being hashed \citep{perlman2016network}. Thus, $\overline{dcba}$, $\overline{cbad}$, $\overline{badc}$, and $\overline{adcb}$ represent the same cyclic symbolic word and are assigned the same identifier and color. Identical normalized words necessarily produce identical hashes; conversely, matching hashes should be verified by direct comparison of the normalized words to exclude possible hash collisions. Equality of the normalized symbolic words indicates equality only under the selected symbolic partition. It does not establish topological equivalence of the corresponding attractors in the full state space.

Symbolic words that do not repeat within the analyzed time window are assigned a normalized Lempel--Ziv (LZ) vocabulary size \citep{pusuluri2018homoclinic,lempel1976complexity}. Gray shading therefore represents the relative complexity of an aperiodic symbolic record under the selected partition and observation window. High LZ complexity is not, by itself, evidence of deterministic chaos or a measure of dynamical instability. Quasiperiodicity, intermittency, long transients, unresolved high-period orbits, numerical limitations, and insufficient observation times may also produce apparently aperiodic symbolic words. Representative neural trajectories can be examined using independent dynamical methods when a specific claim of chaos is required \citep{scully2025chaos}. We therefore describe these regions as aperiodic or complex unless representative trajectories and independent dynamical evidence support a more specific classification.

For the HCO parameter sweeps, we additionally record the mean number of spikes per burst whenever network bursting is detected. For nonbursting trajectories, cell-specific spike counts are used to distinguish tonic spiking, suppression, and quiescence; representative examples are shown in Fig.~\ref{fig:PlantHCOAllTrajectories}. In the \textit{Dendronotus} circuit, spike counts from cells~3 and~4 characterize the state of their HCO-like module under synaptic input from cells~1 and~2. Grids of these two-parameter sweeps reveal how additional synaptic parameters reorganize the network activity regimes, as shown in Figs.~\ref{fig:DendronotusSi3AlphaBetaSweepsgExcVsGInhGridgElectric0.001} and \ref{fig:DendronotusSi3AlphaBetaSweepsgExcVsGInhGridgElectric0.002}. Boundaries in these maps are described as activity-transition boundaries. We use the term \emph{bifurcation} only when the corresponding bifurcation has been established independently.

The methods, including all integration tools and parameters, and source code are available at
\url{https://bitbucket.org/pusuluri_krishna/deterministicchaosprospector/}.

\section{Swim interneuron model}

Following \citet{alacam2015making,scully2025chaos}, we refer to this conductance-based formulation as the swim interneuron (SiN) model. It was developed from the Plant model \cite{plant75,plant76} to specifically describe swim interneurons of the sea slugs \textit{Melibe leonina} and \textit{Dendronotus iris} \citep{bourahmah2024error,scully2024pair,scully2026slow}. Its equations are\begin{eqnarray}
\dot{V} &=& -I_{Na}-I_{K}-I_{Ca}-I_{KCa}-I_{h}-I_{leak}-I_{syn}, \nonumber \\
\dot{Ca} &=& \rho[K_{c}x(V_{Ca}-V+{\rm Ca_{shift}})-Ca], \nonumber \\
\dot{x} &=& ((1/(e^{0.15(-V-50+\rm{x_{shift}})}+1))-x)/\tau_{x},  \nonumber \\
\dot{z} &=& [z_{\infty}(V)-z]/\tau_{z}(V), \quad \mbox{where}~~~z = h, n, y.
\label{equationsPlant}
\end{eqnarray}
Here, $V$ is the membrane voltage and $Ca$ is the intracellular calcium concentration. The variables $h$, $x$, $n$, and $y$ describe sodium-current inactivation, slow activation, potassium-current activation, and $h$-current activation, respectively.    
This is a conductance-based model employing the Hodgkin--Huxley formalism. It includes the fast inward sodium current $I_{Na}$, the outward potassium current $I_{K}$, the slow TTX-resistant calcium current $I_{Ca}$, and the outward calcium-sensitive potassium current $I_{KCa}$. In addition, the model includes the generic ohmic leak current $I_{leak}$, the $h$-current $I_{h}$, and the synaptic current $I_{syn}$ generated by one or more presynaptic neurons. The intracellular calcium concentration $Ca$ is governed by the slowest equation of the model. The complete set of the model equations and their detailed description are presented in the Appendix below.

Compared with the original Plant model of parabolic bursting in the R15 neuron \citep{plant75,plant76,butera1998multirhythmic}, the SiN model introduces two additional bifurcation parameters, ${\rm Ca_{shift}}$ and ${\rm x_{shift}}$. These parameters allow the model to represent swim interneurons that cannot burst in isolation. The parameter ${\rm Ca_{shift}}$ controls the calcium reversal potential, which is known to vary between approximately $80$ and $140~mV$. The parameter ${\rm x_{shift}}$ modifies the dynamics of the slow $x$-variable and can eliminate endogenous bursting due to a hysteresis in the model. When ${\rm Ca_{shift}=x_{shift}}=0$, the system is similar to the original Plant model, which exhibits endogenous bursting. A complete description and detailed bifurcation analysis of the SiN model can be found in \citep{scully2025chaos,scully2024pair}.

\begin{figure}[ht!]
\centering
\includegraphics[width=.85\textwidth]{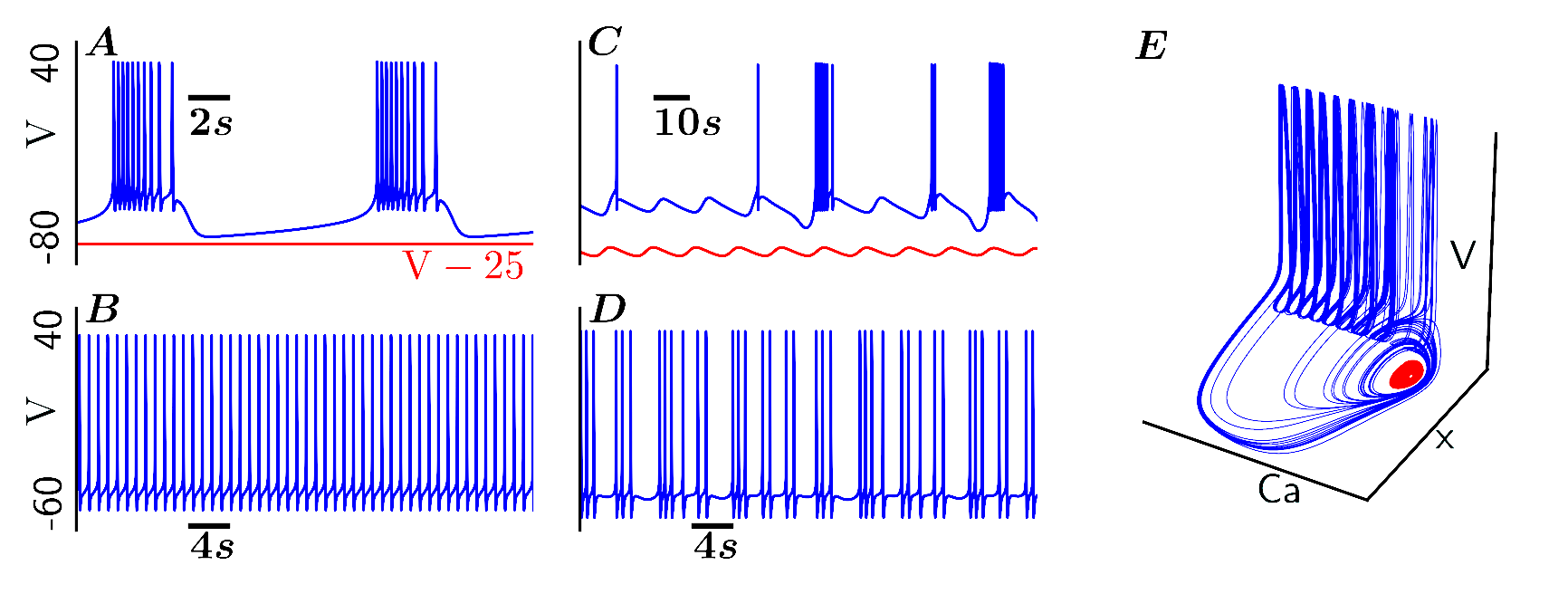}
\caption{Representative activity regimes generated by the SiN model.
(A) Regular bursting with 10 spikes per burst (blue) and a quiescent state
(red). For visual clarity, the quiescent voltage trace is shifted downward by
$25~\mathrm{mV}$.
(B) Tonic-spiking activity.
(C) Bistability between chaotic mixed-mode oscillations (blue) and a
hyperpolarized quiescent state (red). The quiescent voltage trace is shifted
downward by $25~\mathrm{mV}$ for clarity. Trajectories approaching the
quiescent state exhibit damped oscillations, corresponding to spiral
convergence toward the stable equilibrium. The associated phase-space
projection is shown in panel~(E). The basins of the two coexisting attractors
are separated by the stable manifold of a saddle periodic orbit (not shown)
associated with a subcritical Andronov--Hopf bifurcation.
(D) Chaotic activity exhibiting irregular transitions between spiking and
bursting.
(E) Phase-space projection corresponding to the bistable dynamics shown in
panel~(C). }\label{fig:PlantAllTrajectories}
\end{figure}

\subsection{Isolated cell dynamics}

Figure~\ref{fig:PlantSweep} shows a two-parameter sweep of the SiN model, revealing transitions among several activity regimes as the intrinsic cellular parameters are varied. The model exhibits tonic spiking, quiescence, regular bursting with spike-adding transitions, and two distinct forms of chaotic activity. Representative voltage trajectories for these regimes are shown in Fig.~\ref{fig:PlantAllTrajectories}.

Chaotic activity characterized by irregular transitions between spiking and bursting occurs near the boundary between the tonic-spiking and bursting regimes. This region is magnified in Fig.~\ref{fig:PlantSweep}B, with a representative trajectory shown in Fig.~\ref{fig:PlantAllTrajectories}D. A distinct form of chaotic mixed-mode activity occurs near the boundary between the quiescent and bursting regimes, magnified in Fig.~\ref{fig:PlantSweep}C. The representative trajectory in Fig.~\ref{fig:PlantAllTrajectories}C exhibits a variable number of spikes within each burst and a variable number of subthreshold oscillations between successive bursts.

In this parameter region, the chaotic mixed-mode attractor coexists with a hyperpolarized quiescent state. Trajectories approaching the quiescent state exhibit damped oscillations associated with spiral convergence toward the stable equilibrium. The basins of attraction of the coexisting states are separated by the stable manifold of a saddle periodic orbit associated with a subcritical Andronov--Hopf bifurcation.

\subsection{Emergent HCO bursting from non-endogenous bursters}

\begin{figure}[ht!]
\centering
\includegraphics[width=0.6\textwidth]{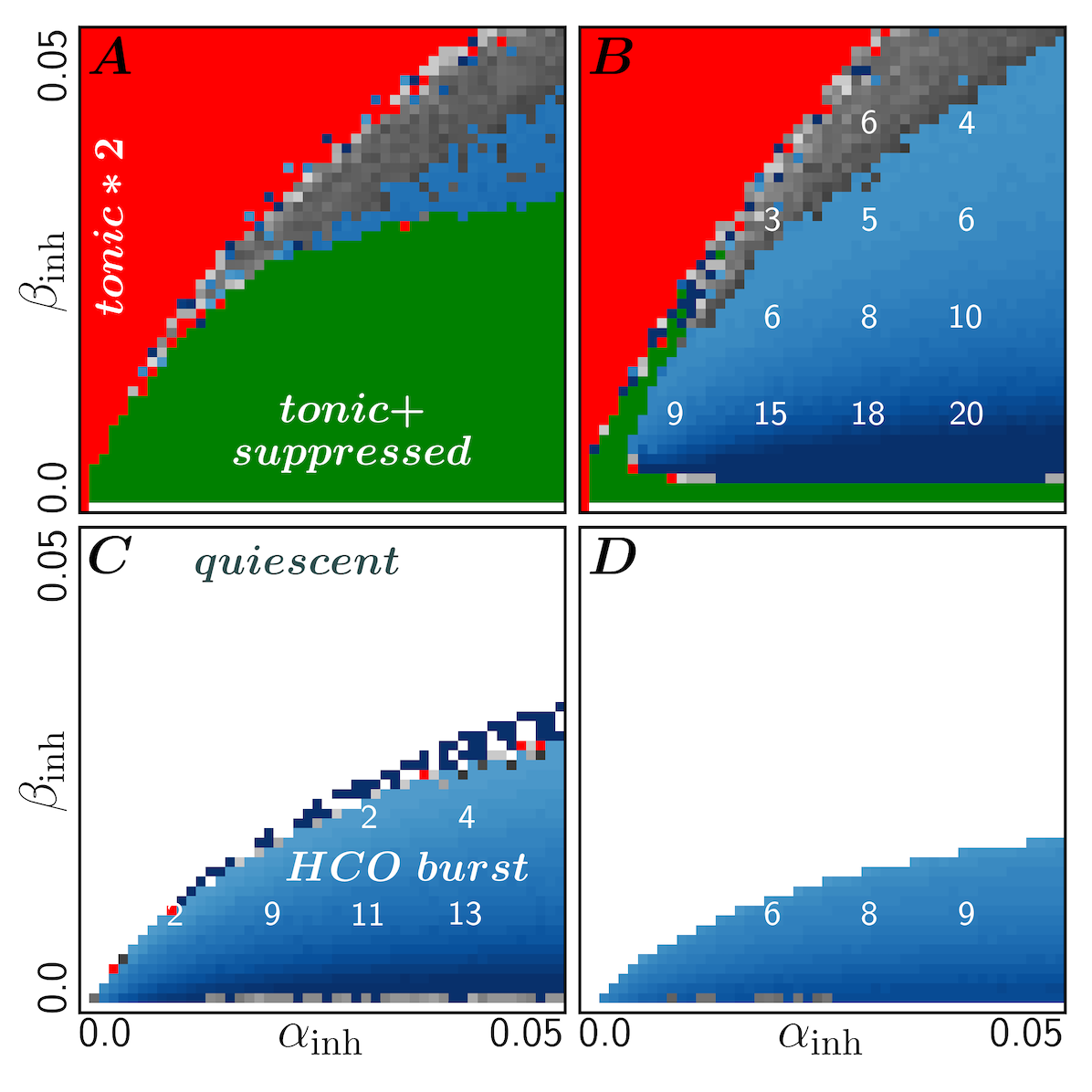}
\caption{Parameter regions supporting emergent network bursting in a half-center
oscillator (HCO) across the synaptic
$\left(\alpha_{\mathrm{inh}},\,\beta_{\mathrm{inh}}\right)$ parameter plane.
The HCO consists of two identical, mutually inhibitory SiN model neurons
(Fig.~\ref{fig:PlantHCOSymbolsNetwork}A) that do not burst intrinsically.
The intrinsic parameter sets used in panels~(A)--(D) correspond to the four
black dots in Fig.~\ref{fig:PlantSweep}, located near the boundary between
tonic spiking and quiescence. In all four panels,
${\mathrm{x}}_{\mathrm{shift}}=-2.0$, while
${\mathrm{Ca}}_{\mathrm{shift}}=-55$, $-50$, $-45$, and
$-40~\mathrm{mV}$ in panels~(A)--(D), respectively.
For each intrinsic parameter set, a two-parameter sweep of the slow
mutual-inhibitory synaptic parameters
$\alpha_{\mathrm{inh}}$ and $\beta_{\mathrm{inh}}$ reveals how intrinsic
cellular excitability and synaptic dynamics jointly regulate the HCO
behavior.
The identified regimes include tonic spiking by both cells (red),
quiescence of both cells (white), chaotic bursting (gray), a
winner-take-all state in which one tonically spiking cell suppresses the
other into quiescence (green), and emergent anti-phase HCO bursting (blue).
Within the blue regions, the color intensity represents the average number
of spikes per burst in cell~2, with darker shades indicating more spikes per
burst. In the absence of synaptic coupling, the neurons are intrinsic tonic
spikers for the parameter sets in panels~(A) and~(B), whereas they are
quiescent for those in panels~(C) and~(D). These intrinsic regimes are
reflected by the red and white regions, respectively, at low values of
$\alpha_{\mathrm{inh}}$.
As the intrinsic parameters
${\mathrm{x}}_{\mathrm{shift}}$ and
${\mathrm{Ca}}_{\mathrm{shift}}$ move farther from the transition boundaries
among tonic spiking, quiescence, and intrinsic bursting in
Fig.~\ref{fig:PlantSweep}, the blue regions supporting emergent HCO bursting
shrink and eventually disappear, as illustrated in the right panel of
Fig.~\ref{fig:DendronotusSi3CaShiftvsXShift1_Si3HCO}.}\label{fig:PlantHCOGrid}
\end{figure}

\begin{figure}[ht!]
\centering
\includegraphics[width=.7\textwidth]{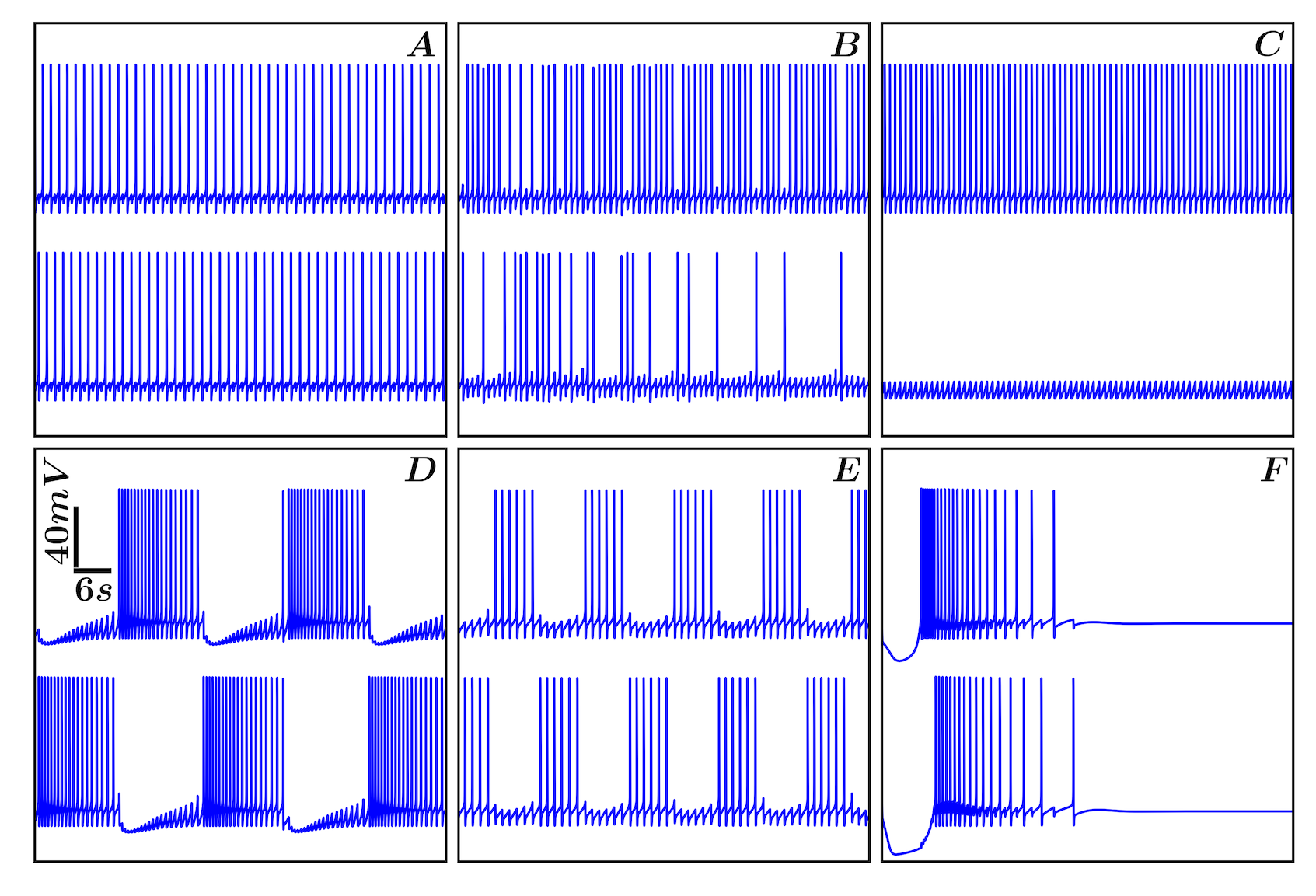}
\caption{Representative voltage trajectories and emergent activity patterns in the
half-center oscillator (HCO) presented in Fig.~\ref{fig:PlantHCOSymbolsNetwork}A. The HCO consists of two reciprocally inhibitory cells that do not burst intrinsically.
(A) Both cells exhibit tonic spiking.
(B) The HCO produces irregular chaotic bursting.
(C) A winner-take-all state in which one cell fires tonically and suppresses
the other cell into quiescence.
(D) Emergent periodic bursting with ~20 spikes per burst.
(E) Emergent periodic bursting with ~6 spikes per burst.
(F) Following an external perturbation that evokes a brief episode of
spiking, both cells return to stable quiescence. Because the two cells and their reciprocal synaptic connections are symmetric, the winner-take-all regime in panel~(C) is bistable. Depending on
the initial conditions or applied perturbations, either cell~1 can fire
tonically and suppress cell~2, as shown, or cell~2 can fire tonically and
suppress cell~1. Panels~(A)--(C) correspond to representative parameter values from
Fig.~\ref{fig:PlantHCOGrid}A with
${\rm Ca}_{\rm shift}=-55~\mathrm{mV}$:
(A) $\alpha=0.01$ and $\beta=0.04$;
(B) $\alpha=0.035$ and $\beta=0.04$; and
(C) $\alpha=0.02$ and $\beta=0.01$.
Panels~(D) and~(E) are sampled from Fig.~\ref{fig:PlantHCOGrid}B with
${\rm Ca}_{\rm shift}=-50~\mathrm{mV}$:
(D) $\alpha=0.04$ and $\beta=0.01$; and
(E) $\alpha=0.04$ and $\beta=0.03$.
Panel~(F) is sampled from Fig.~\ref{fig:PlantHCOGrid}C with
${\rm Ca}_{\rm shift}=-45~\mathrm{mV}$,
$\alpha=0.01$, and $\beta=0.04$.
 }\label{fig:PlantHCOAllTrajectories}
\end{figure}

In this section, we discuss the emergence of HCO bursting in a pair of identical cells that do not burst intrinsically, given by Eq.~(\ref{equationsPlant}), and mutually coupled through slow inhibitory synapses described by the following alpha-synapse model \cite{scully2024pair}:
\begin{eqnarray}
I_{syn} &=& g_{inh}S(V_{post}-V_{rev}), \nonumber \\
\dot{S} &=& \alpha(1-S)/(1+e^{-10(V_{pre}-V_{thr})}) - \beta S.
\label{equationsSynPlant}
\end{eqnarray}
Here, $V_{pre}$ and $V_{post}$ denote the voltages of the pre- and post-synaptic cells, respectively, and $V_{rev}$ is the reversal potential of the synaptic current set here to $-70mV$ for inhibitory synapses, while $V_{thr}=-20mV$ defines the synaptic threshold. The parameters $g_{inh}=0.01$, $\alpha$ and $\beta$ are identical for both synapses in the HCO, with particular values specified in the figure captions. In this phenomenological alpha-synapse model, $\alpha$ determines the activation rate when the presynaptic voltage exceeds the synaptic threshold, whereas $\beta$ determines the decay rate. Increasing $\alpha$ or decreasing $\beta$ increases the magnitude and duration of the synaptic gating variable $S(t)$ and thereby increases the effective inhibitory influence at fixed $g_{inh}$. These parameters characterize synaptic kinetics and should not be interpreted literally as measures of neurotransmitter amount or clearance. Their dynamical effects can be tested experimentally using conductance waveforms implemented by dynamic clamp.

Figure~\ref{fig:PlantHCOGrid} shows four bi-parametric sweeps for the HCO (Fig.~\ref{fig:PlantHCOSymbolsNetwork}A), composed either of two intrinsically tonic-firing cells in panels A and B or two quiescent cells in panels C and D, using parameters ${\rm x_{shift}}=-2.$ and ${\rm Ca_{shift}}= [-55., -50, -45, -40]$ sampled at the black dots in Fig.~\ref{fig:PlantSweep} near the boundary between tonic spiking and quiescence. As we vary the slow synaptic properties $\alpha_{inh}$ and $\beta_{inh}$ of mutual inhibition in the biparametric sweeps, the network exhibits a broad range of activity patterns, including tonic spiking of both cells (red regions), quiescence of both cells (white), chaotic bursting (gray), suppression of one cell into quiescence by the other tonically firing cell (green), and emergent HCO bursting (blue). Within the blue regions, the average number of spikes per burst in cell $2$ is also indicated, with darker blue corresponding to a larger number of spikes per burst. The red tonic-spiking regions at low values of $\alpha_{inh}$ in panels A and B, and the quiescent regions (white) in panels~C and D, are consistent with the intrinsic behaviors of the cells at the corresponding ${\rm Ca_{shift}}$ values.

The corresponding voltage trajectories are shown in Figs.~\ref{fig:PlantHCOAllTrajectories}A,B and C for the regimes of tonic spiking in both cells, chaotic bursting, and suppression of one cell into quiescence by the other tonic-spiking cell, respectively, at parameter values sampled from Fig.~\ref{fig:PlantHCOGrid}A. When cell~2 fires tonically and suppresses cell~1, the symmetry of the network implies the existence of a symmetry-related state in which cell~1 fires tonically and suppresses cell~2. Thus, the system is bistable at these parameter values, with the realized state determined by the initial conditions or external perturbations. Representative examples of emergent bursting with 20 spikes per burst and 6 spikes per burst are shown in Figs.~\ref{fig:PlantHCOAllTrajectories}D and E for parameters sampled from Fig.~\ref{fig:PlantHCOGrid}B. Figure~\ref{fig:PlantHCOAllTrajectories}F shows the stable quiescent state of both cells, sampled from the white region of Fig.~\ref{fig:PlantHCOGrid}C. An external perturbation can briefly induce spiking in these cells, but the trajectories eventually return to the stable quiescent state.

As the synapses are strengthened in Fig.\ref{fig:PlantHCOGrid} by increasing values of $\alpha_{inh}$, or decreasing values of $\beta_{inh}$, the network gradually transitions from tonic spiking through chaotic bursting and emergent network bursting to suppression in the case of intrinsic tonic spikers (Figs.~\ref{fig:PlantHCOGrid}A and B). For intrinsic quiescent cells (Figs.\ref{fig:PlantHCOGrid}~C and D), the network transitions from quiescence through chaotic bursting to emergent network bursting as the synapse is gradually strengthened. If ${\rm x_{shift}}$ and ${\rm Ca_{shift}}$ values are sampled further away from the boundary between tonic spiking and quiescence (as well as bursting) in Fig.~\ref{fig:PlantSweep}, the regions of emergent HCO bursting (blue) shrink and eventually vanish, eliminating network bursts (compare with Fig.~\ref{fig:DendronotusSi3CaShiftvsXShift1_Si3HCO}B). The parametric sweeps thus reveal how the synergistic interactions between intrinsic and synaptic properties control the emergence of HCO bursting, as well as quantitative properties of the resulting rhythm, such as the average number of spikes per burst.

\subsection{Slow--fast release mechanism of emergent HCO-bursting}

\begin{figure}[t!]
\centering
\includegraphics[width=0.8\textwidth]{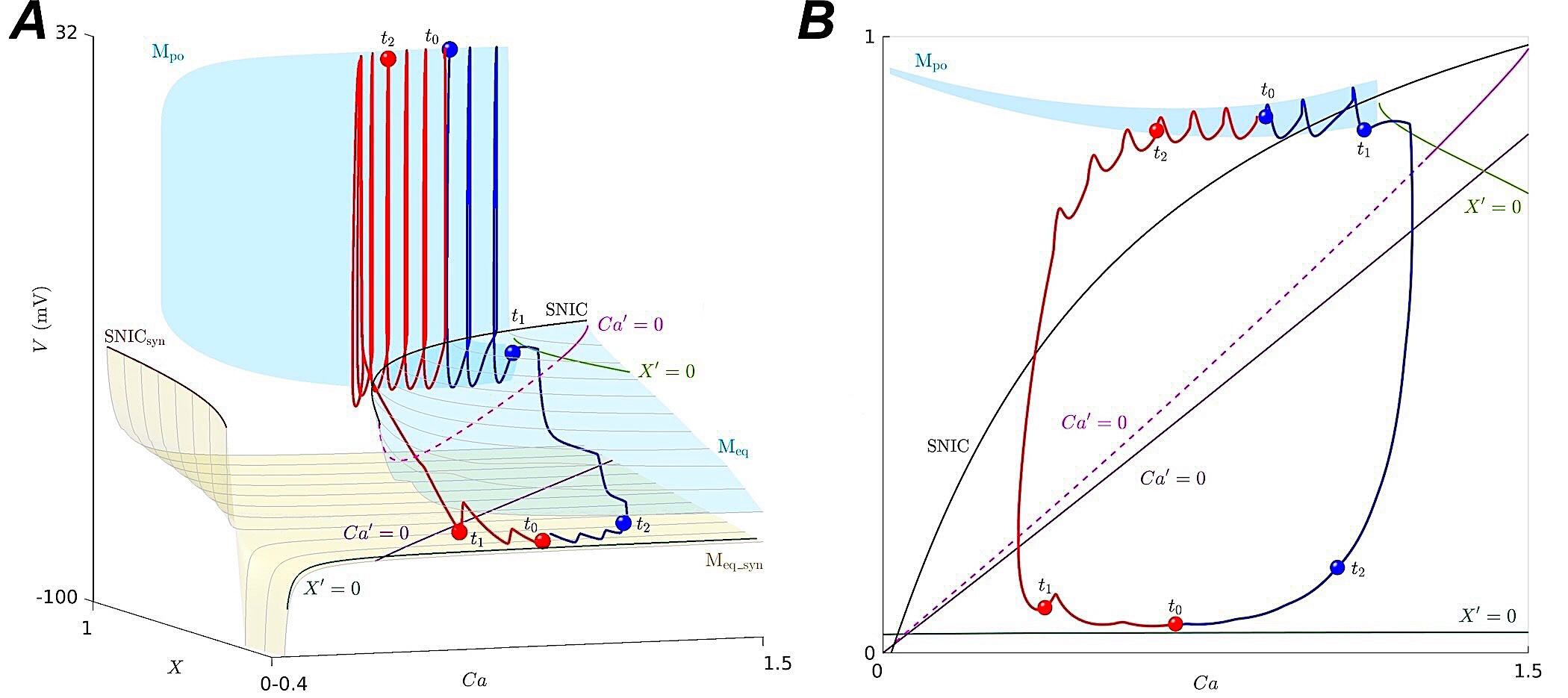}
\caption{Slow--fast geometry underlying release-mediated bursting in an HCO composed
of two SiN model neurons.
(A) Three-dimensional
$\left(V,X,\mathrm{Ca}\right)$ projection of the HCO trajectory, with the
orbits of the two neurons shown in red and blue. The projection also shows
the one-dimensional equilibrium slow manifold
$\mathcal{M}_{\mathrm{eq}}^{}$, its synaptically inhibited counterpart
$\mathcal{M}_{\mathrm{eq\text{-}syn}}^{}$, and the two-dimensional
tonic-spiking manifold $\mathcal{M}_{\mathrm{po}}^{}$. The intersections
of the $\dot{\mathrm{Ca}}=0$ and $\dot{X}=0$ slow nullclines with the
equilibrium manifolds, together with the
$\mathrm{SNIC}$ and $\mathrm{SNIC}_{\mathrm{syn}}$ bifurcation curves,
organize the slow evolution of the network trajectory.
Beginning at $t_0$, the blue neuron approaches the stable inhibited
equilibrium and becomes silent at $t_1$, thereby releasing the red neuron
from inhibition. The red neuron subsequently approaches the
tonic-spiking manifold $\mathcal{M}_{\mathrm{po}}^{}$ near $t_2$, resumes
spiking, and inhibits the blue neuron. Alternation of these release events
generates the anti-phase bursting rhythm of the HCO.
(B) Corresponding
$\left(\mathrm{Ca},X\right)$ projection of the SiN model phase space,
illustrating the network hysteresis underlying the HCO rhythm. A
representative HCO trajectory is superimposed to show the progression of
the instantaneous states of the two neurons through the silent and active
phases of the bursting cycle. Further details on the roles of the slow
manifolds, nullclines, and bifurcation curves in organizing the SiN model
dynamics are provided in Ref.~\citep{scully2024pair}.}
\label{fig:plant_release}
\end{figure}

The parameter sweeps identify regions in which mutual inhibition recruits nonbursting SiN-model neurons into alternating bursts. Figure~\ref{fig:plant_release} provides a complementary geometrical interpretation of the mechanism underlying this pairwise rhythm. In a slow--fast decomposition, the active neuron evolves near the tonic-spiking manifold $\mathrm{M}{\rm po}$, whereas inhibition drives its partner toward the stable equilibrium manifold $\mathrm{M}{\rm eq-syn}$. As the active cell approaches the termination region of the spiking manifold, the inhibited cell is released. The released cell then returns toward $\mathrm{M}{\rm po}$ and begins spiking, thereby inhibiting its partner and exchanging the active and silent roles. The intersections of the $Ca'=0$ and $X'=0$ slow nullclines with the uninhibited and inhibited equilibrium manifolds organize the slow drift, while the SNIC and $\mathrm{SNIC}{\rm syn}$ curves mark the loss of the corresponding fast-subsystem equilibria.
This construction supports a release mechanism for the representative HCO orbit shown: alternation arises because inhibition stabilizes the silent branch until activity in the presynaptic neuron terminates and the inhibited cell is released. It does not imply that every bursting region in Fig.~\ref{fig:PlantHCOGrid} is generated by an identical mechanism; establishing such a stronger statement would require continuation of the relevant invariant sets across the full parameter plane. Rather, the purpose of this construction is to explain how mutual inhibition can generate HCO-emergent bursting in a pair of cells that do not burst intrinsically. The next section provides a stronger hierarchical test by considering a \textit{Dendronotus} parametrization for which the isolated HCO itself no longer bursts.

\section{Emergent bursting in the 4-cell \textit{Dendronotus} swim CPG model}

\begin{figure}[ht!]
\centering
\includegraphics[width=0.75\textwidth]{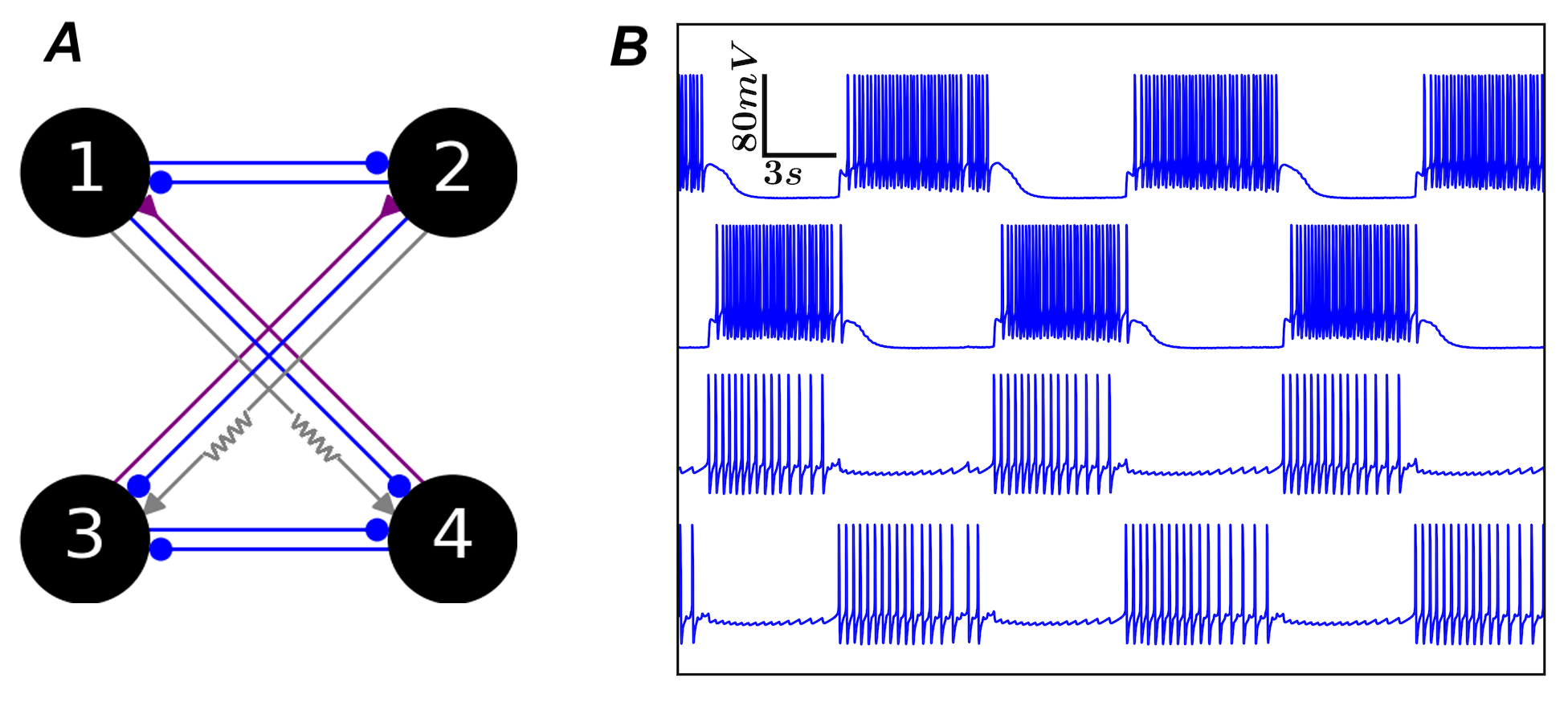}  
\caption{Architecture and network-generated bursting in a simplified four-cell circuit
modeled after the \textit{Dendronotus} swim central pattern generator (CPG).
(A) Schematic of the network, which consists of two interacting half-center
oscillators (HCOs) whose synaptic interactions act synergistically to produce
the collective rhythm. Inhibitory chemical and excitatory synapses, and rectifying electrical connections are shown in blue, purple,
and gray, respectively.
(B) Representative bursting rhythm generated by the coupled four-cell
network. The activity pattern resembles characteristic neurophysiological
recordings from the animal, although none of the individual model cells
bursts intrinsically. In the absence of synaptic coupling, cells~1 and~2 are
quiescent, whereas cells~3 and~4 are tonic spikers. Thus, the bursting rhythm
is an emergent property of the coupled network rather than an intrinsic
property of its component neurons.
The parameter values used to generate the rhythm in panel~B are
$g_{14\_23\_{\mathrm{inh}}}=0.006$,
$g_{41\_32\_{\mathrm{exc}}}=0.04$,
$\alpha_{34\_43\_{\mathrm{inh}}}=0.04$, and
$\beta_{34\_43\_{\mathrm{inh}}}=0.006$.
This parameter set corresponds to the regime with 18 spikes per burst shown
in Fig.~\ref{fig:DendronotusSi3AlphaBetaSweepsgExcVsGInhGridgElectric0.002}D.}\label{fig:DendronotusNetworkTrajectory}
\end{figure}

The CPG governing the swimming behavior of the sea slug \textit{Dendronotus iris} is described in \citep{katz2017neural,sakurai2016central,sakurai2017artificial,scully2026slow}. The simplified 4-cell circuit and the bursting rhythm generated by the network model, which resembles the rhythmic activity observed in the animal, are presented in Fig.~\ref{fig:DendronotusNetworkTrajectory}. These network bursting oscillations underlie the rhythmic movements of the animal during swimming. The circuit essentially consists of a pair of interacting HCOs whose connections act synergistically. Cells $1$ and $2$ mutually inhibit each other (blue connections), as do cells $3$ and $4$. In addition, strong cross excitation (purple) projects from $3$ to $2$ and from $4$ to $1$, while inhibitory connections project in the opposite directions (blue). A pair of electrical cross connections also exists between these cells. Recent recordings by A. Sakurai suggest that these connections may correspond to unidirectional rectified electrical coupling. In the absence of synaptic coupling, none of the cells burst intrinsically. Cells $1$ and $2$ are quiescent, whereas cells $3$ and $4$ are modeled as intrinsic tonic spikers. In the remainder of this section, we use parametric sweeps to examine how the synergistic interactions among different network parameters generate emergent bursting and regulate important features of the resulting rhythm, including the number of spikes per burst and burst duration, as illustrated in Fig.~\ref{fig:DendronotusNetworkTrajectory} and Fig.~\ref{fig:DendronotusAllTrajectories}C (with 18 or 14 spikes per burst in cell $4$, here).
The model employed for this network is similar to the SiN model in Eq.~(\ref{equationsPlant}), although several properties of the ionic and synaptic currents are adjusted to better reproduce experimental recordings, including shorter action potentials, faster sodium and potassium currents, and rescaled voltage-dependent time constants. Detailed equations and parameters of the model are available in the software repository accompanying this study: \ \url{https://bitbucket.org/pusuluri\_krishna/deterministicchaosprospector/}. \

Inhibitory and excitatory connections are modeled using calibrated slow $\alpha$-synapses following \citep{scully2025chaos} as follows:
\begin{eqnarray}
I_{syn} &=& g_{inh/exc}S(V_{post}-V_{rev}), \nonumber \\
\dot{S} &=& \alpha_{inh/exc}(1-S)/(1+e^{-10(V_{pre}-V_{thr})}) - \beta_{inh/exc} S.
\label{equationsSynDendronotus}
\end{eqnarray}
with default values of the synaptic connections such as $V_{rev}=-70mV$ and $V_{thr}=-20mV$ for the inhibitory reversal potential and the synaptic threshold, respectively, unless otherwise specified. The network is bilaterally symmetric from left to right. Accordingly, $g_{12\_inh}=g_{21_inh}=g_{12\_21\_inh}=0.02$ , represents the mutual inhibitory strength between cells $1$ and $2$, while the corresponding parameters controlling the synapse are given by $\alpha_{12\_21\_inh}=0.04$ and $\beta_{12\_21\_inh}=0.001$. The mutual inhibition between cells $3$ and $4$ is given by $g_{34\_43\_inh}=0.04$, $\alpha_{34\_43\_inh}=0.01$, $\beta_{34\_43\_inh}=0.005$; cross excitatory coupling by $g_{41\_32\_exc}=0.08$, $\alpha_{41\_32\_exc}=0.02$, $\beta_{41\_32\_exc}=0.004$; cross inhibition by $g_{14\_23\_inh}=0.0055$, $\alpha_{14\_23\_inh}=0.01$, $\beta_{14\_23\_inh}=0.001$; and rectified electrical coupling by $g_{14\_23\_elec}=0.002$.

\subsection{Absence of bursting in isolated cells and HCOs}

\begin{figure}[t!]
\centering
\includegraphics[width=.7\textwidth]{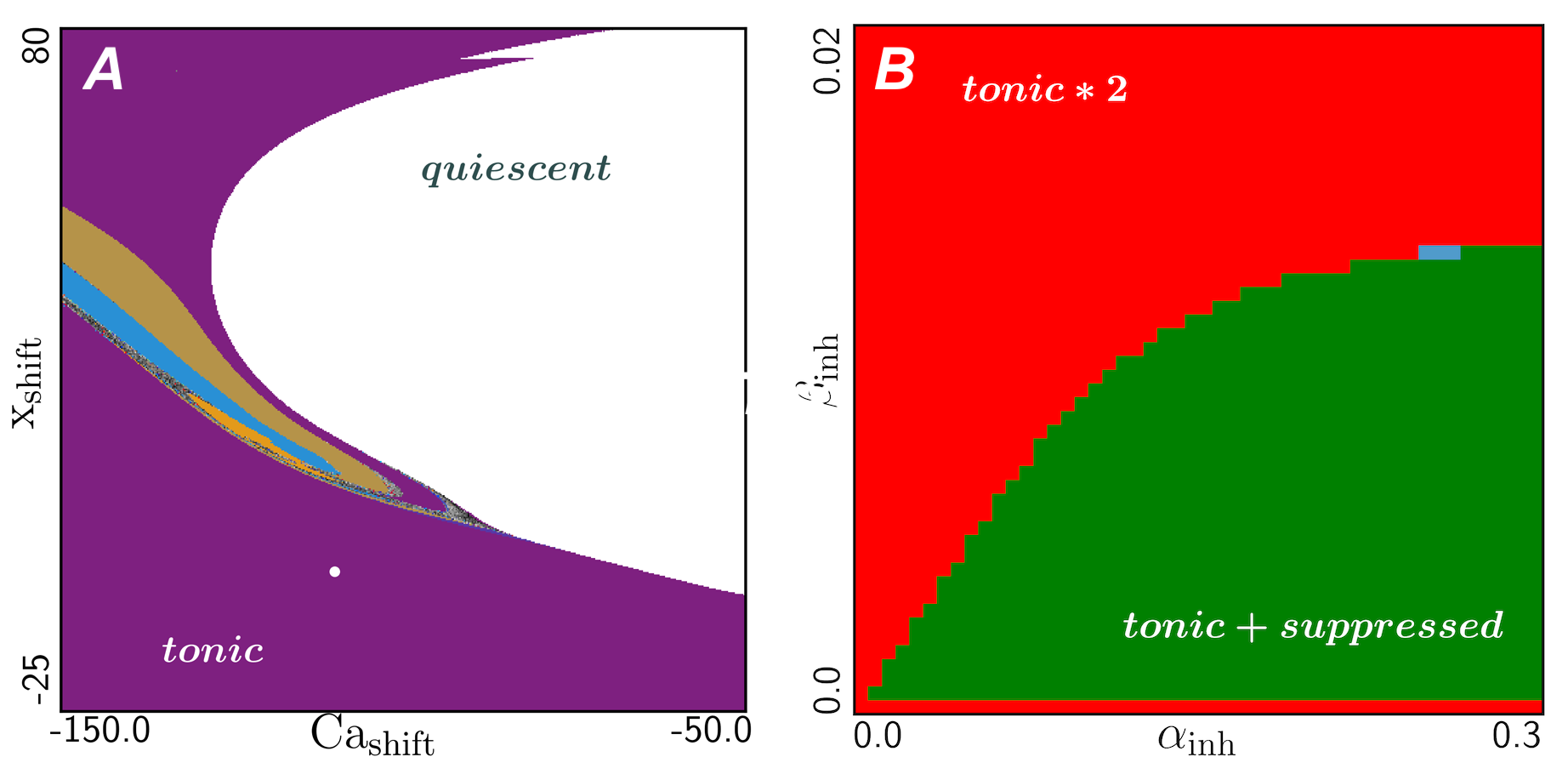}
\caption{(A) Two-parameter sweep of the modified SiN model in the
$\left(\mathrm{Ca}_{\mathrm{shift}},
\,\mathrm{x}_{\mathrm{shift}}\right)$ parameter plane. Across most of the
examined region, the isolated model neuron exhibits either tonic-spiking
activity (purple) or quiescence (white), while intrinsic bursting occurs only
within comparatively narrow parameter regions. This organization differs
from that of the original SiN model shown in
Fig.~\ref{fig:PlantSweep}. The white dot at
$\mathrm{Ca}_{\mathrm{shift}}=-110$ and
$\mathrm{x}_{\mathrm{shift}}=-3.5$, located within the tonic-spiking region,
marks the intrinsic parameter values assigned to cells~3 and~4 of the
network shown in Fig.~\ref{fig:DendronotusNetworkTrajectory}.
(B) Two-parameter sweep of the HCO formed by cells~3 and~4 as the
mutual-inhibitory synaptic parameters
$\alpha_{34\_43\_{\mathrm{inh}}}$ and
$\beta_{34\_43\_{\mathrm{inh}}}$ are varied. In the absence of synaptic
connections with cells~1 and~2, the isolated HCO produces only tonic-spiking
or suppressed states throughout the examined parameter region; no emergent
bursting is observed. Thus, reciprocal inhibition between cells~3 and~4
alone is insufficient to generate bursting within the parameter ranges
considered.}
\label{fig:DendronotusSi3CaShiftvsXShift1_Si3HCO}
\end{figure}

Figure~\ref{fig:DendronotusSi3CaShiftvsXShift1_Si3HCO}A shows the ${\rm Ca_{shift}}$ vs. ${\rm x_{shift}}$ parametric sweep of the SiN-neuron model. In isolation, these neurons predominantly exhibit tonic spiking or quiescence, with intrinsic bursting confined to a comparatively smaller region of parameter space than in the SiN model shown in Fig.~\ref{fig:PlantSweep}. The parameters ${\rm Ca_{shift}}=-110.$, ${\rm x_{shift}}=-3.5$ in the tonic spiking region (white dot) are chosen to model the intrinsically tonic-spiking cells $3$ and $4$ in the network of Fig.\ref{fig:DendronotusNetworkTrajectory}. We then analyze the HCO formed by these cells by varying the synaptic parameters $\alpha_{34\_43\_inh}$ and $\beta_{34\_43\_inh}$ in Fig.\ref{fig:DendronotusSi3CaShiftvsXShift1_Si3HCO}B. This sweep shows that, in the absence of synaptic connections with cells $1$ and $2$, the isolated HCO produces only tonic spiking or suppression and does not exhibit emergent bursting. This contrasts with the HCO considered in Fig.~\ref{fig:PlantHCOGrid}, where mutual inhibition between a different pair of SiN model neurons is sufficient to generate emergent bursting.

\subsection{Synergistic excitation, inhibition, and electrical coupling generate emergent 4-cell bursting in the swim CPG model}

\begin{figure}[ht!]
\centering
\includegraphics[width=.8\textwidth]{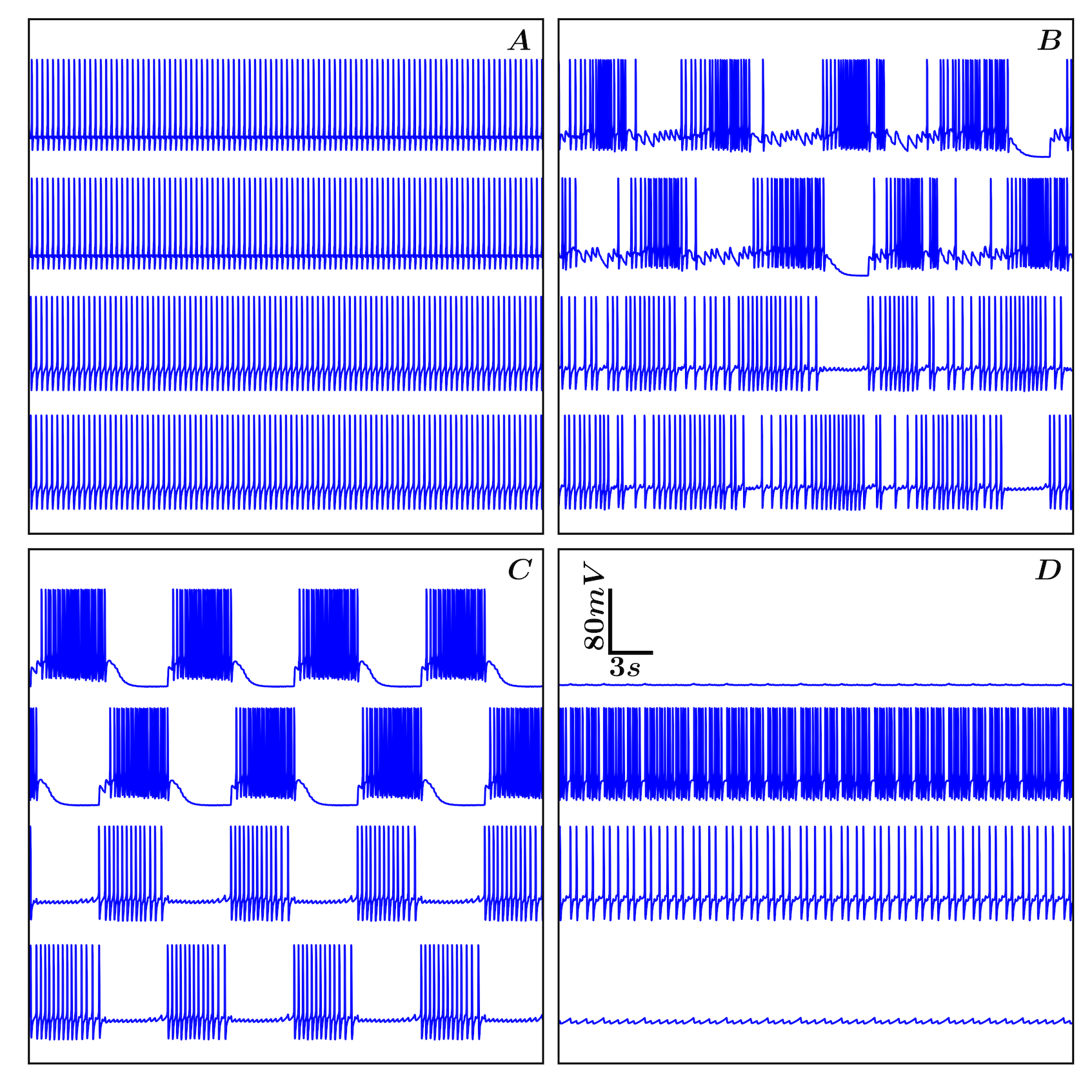}
\caption{Representative activity regimes observed in the four-cell network as the
slow mutual-inhibitory synaptic parameters are varied.
(A) All four cells exhibit tonic-spiking activity under weak mutual
inhibition.
(B) The network displays irregular, weakly developed bursting near the
boundary between tonic spiking and organized bursting.
(C) The network generates pronounced emergent bursting with approximately
14 spikes per burst.
(D) A winner-take-all regime in which cells~1 and~4 are suppressed by
inhibition from the tonically active cells~2 and~3.
The cross-excitatory and cross-inhibitory conductances are fixed at
$g_{41\_32\_{\mathrm{exc}}}=0.04$ and
$g_{14\_23\_{\mathrm{inh}}}=0.006$, respectively; see
Fig.~\ref{fig:DendronotusSi3AlphaBetaSweepsgExcVsGInhGridgElectric0.002}D.
For panels~A--C,
$\beta_{34\_43\_{\mathrm{inh}}}=0.005$, while
$\alpha_{34\_43\_{\mathrm{inh}}}=0.004$, $0.01$, and $0.02$,
respectively. For panel~(D),
$\alpha_{34\_43\_{\mathrm{inh}}}=0.04$ and
$\beta_{34\_43\_{\mathrm{inh}}}=0.004$. }\label{fig:DendronotusAllTrajectories}
\end{figure}

\begin{figure}[ht!]
\centering
\includegraphics[width=0.9\textwidth]{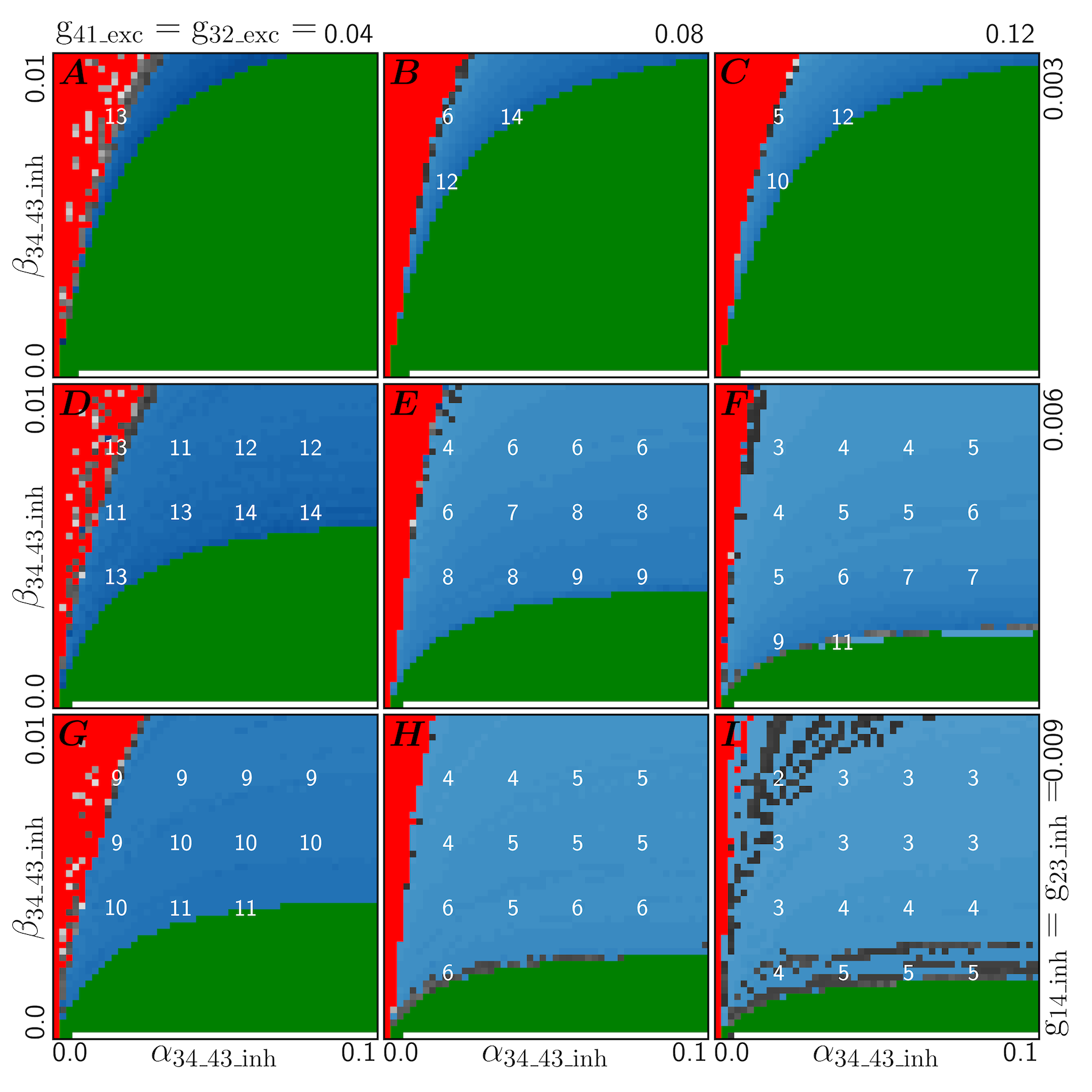}
\caption{3x3 grid of bi-parametric sweeps in the synaptic 
$\left(\alpha_{34\_43\_{\mathrm{inh}}},
\,\beta_{34\_43\_{\mathrm{inh}}}\right)$-parameter plane. The cross-excitatory
conductance $g_{41\_32\_{\mathrm{exc}}}$ varies horizontally across columns,
whereas the cross-inhibitory conductance
$g_{14\_23\_{\mathrm{inh}}}$ varies vertically across rows. Each panel shows
how the network dynamics change as the slow synaptic parameters governing
mutual inhibition between the intrinsically tonic-spiking cells~3 and~4 are
varied.
The sweeps demonstrate that interactions among the mutual-inhibitory,
cross-excitatory, and cross-inhibitory synapses produce several distinct
network regimes, including tonic spiking (red), chaotic bursting (gray),
suppression (green), and emergent network bursting (wide blue regions) with indicated numbers of spikes per burst. Within the bursting
regions, the displayed numbers indicate the average number of spikes per
burst in cell~4. Representative voltage trajectories for these activity
regimes are shown in Fig.~\ref{fig:DendronotusAllTrajectories}.
The electrical coupling conductance is fixed at
$g_{14\_23\_{\mathrm{elec}}}=0.001$. The corresponding grid for the stronger
electrical coupling
$g_{14\_23\_{\mathrm{elec}}}=0.002$ is shown in
Fig.~\ref{fig:DendronotusSi3AlphaBetaSweepsgExcVsGInhGridgElectric0.002}.}\label{fig:DendronotusSi3AlphaBetaSweepsgExcVsGInhGridgElectric0.001}
\end{figure}
\begin{figure}[ht!]
\centering
\includegraphics[width=0.9\textwidth]{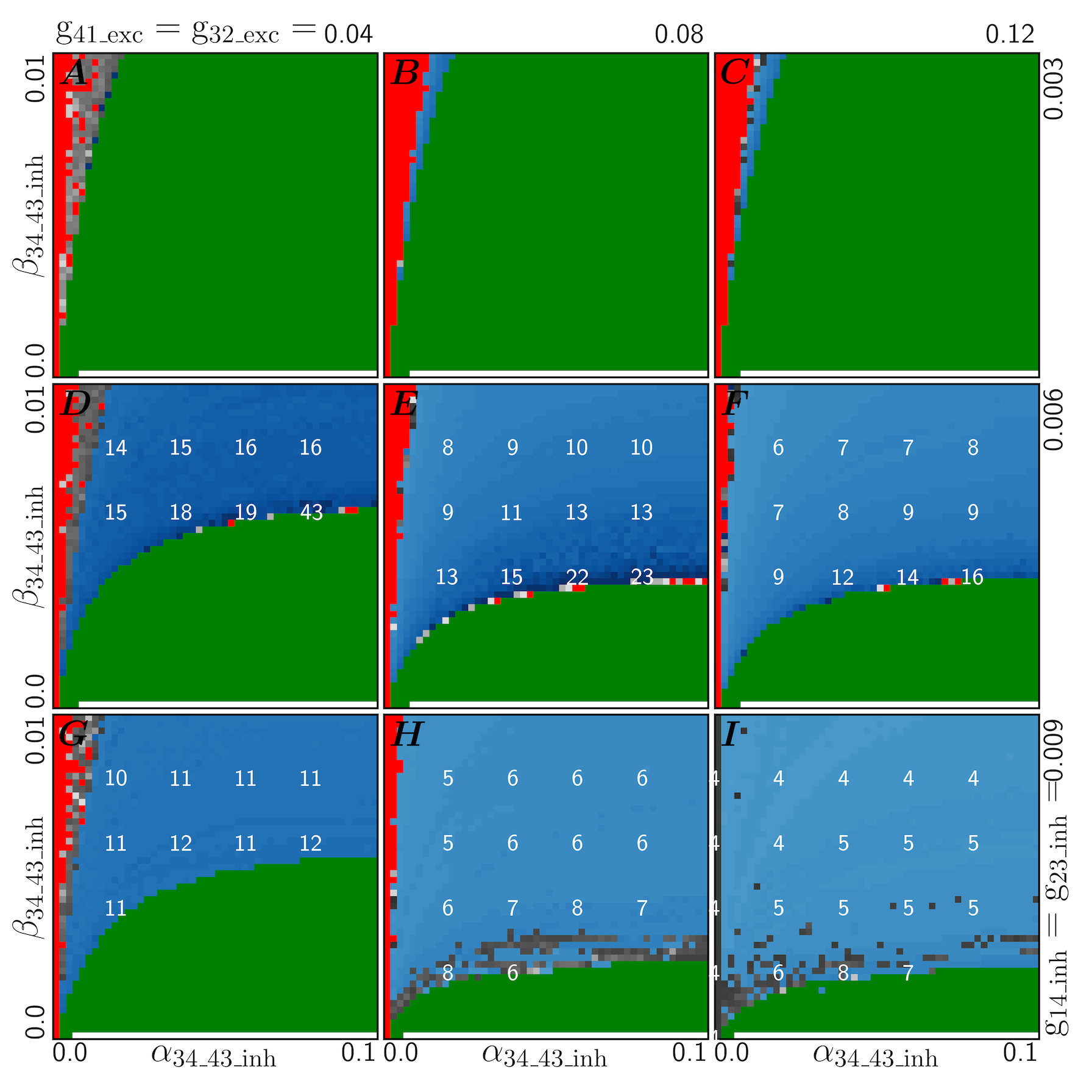}
\caption{3x3 grid of bi-parametric sweeps in the synaptic
$\left(\alpha_{34\_43\_{\mathrm{inh}}},
\,\beta_{34\_43\_{\mathrm{inh}}}\right)$-parameter plane for an electrical coupling
conductance of $g_{14\_23\_{\mathrm{elec}}}=0.002$. The cross-excitatory
conductance $g_{41\_32\_{\mathrm{exc}}}$ varies horizontally across columns,
whereas the cross-inhibitory conductance
$g_{14\_23\_{\mathrm{inh}}}$ varies vertically across rows. Each panel shows
the network activity produced as the slow mutual-inhibitory synaptic
parameters between cells~3 and~4 are varied. The notation, color coding, and
displayed spike counts follow
Fig.~\ref{fig:DendronotusSi3AlphaBetaSweepsgExcVsGInhGridgElectric0.001},
which presents the corresponding sweeps for the weaker electrical coupling
$g_{14\_23\_{\mathrm{elec}}}=0.001$.}
\label{fig:DendronotusSi3AlphaBetaSweepsgExcVsGInhGridgElectric0.002}
\end{figure}

For modeling the full 4-cell circuit, the intrinsic parameters ${\rm Ca_{shift}}$ and ${\rm x_{shift}}$ for the tonic-spiking cells $3$ and $4$ are set to ${\rm Ca_{shift}}=-110.$, ${\rm x_{shift}}=-3.5$ (white dot in Fig.~\ref{fig:DendronotusSi3CaShiftvsXShift1_Si3HCO}A), whereas those for the intrinsically quiescent cells $1$ and $2$ are set to ${\rm Ca_{shift}}=0.$, ${\rm x_{shift}}=-3.5$ (in the white quiescent region to the right, outside the plot boundaries of Fig.~\ref{fig:DendronotusSi3CaShiftvsXShift1_Si3HCO}A). In the absence of cross inhibition, cross excitation, and electrical coupling between the two HCOs, neither HCO is capable of emergent bursting. The parametric sweep for the tonic-spiking HCO formed by cells $3$,$4$ in Fig.~\ref{fig:DendronotusSi3CaShiftvsXShift1_Si3HCO}B shows only tonic spiking and suppression, while the corresponding sweep for the HCO formed by cells $1$,$2$ would contain only the white quiescent region. Thus, neither isolated HCO generates the bursting rhythm observed in the full circuit. In the following sections, we keep the synaptic properties of the quiescent HCO formed by cells $1$,$2$ fixed and examine how the dynamics of the HCO formed by the tonic-spiking cells $3$, $4$, and consequently those of the full network, are shaped by the synergistic action of cross excitation, cross inhibition, and electrical coupling, together with the synaptic properties of the mutual inhibition between cells $3$ and $4$.

The voltage trajectories corresponding to the different activity regimes observed in the 4-cell network are shown in Fig.~\ref{fig:DendronotusAllTrajectories}. At parameter values given by $g_{41\_32\_exc}=0.04$, $g_{14\_23\_inh}=0.006$ (see Fig.~\ref{fig:DendronotusSi3AlphaBetaSweepsgExcVsGInhGridgElectric0.002}D), and $\beta_{34\_43\_inh}=0.005$, the network exhibits several distinct dynamical regimes as the mutual inhibition between cells $3$,$4$, regulated by $\alpha_{34\_43\_inh}$, is gradually increased through $0.004, 0.01, 0.02$. These include tonic spiking of all cells, as shown in Fig.\ref{fig:DendronotusAllTrajectories}A, chaotic spiking/bursting in Fig.\ref{fig:DendronotusAllTrajectories}B, and emergent network bursting with 14 spikes per burst in cell $4$, shown in Fig.\ref{fig:DendronotusAllTrajectories}C. At $\alpha_{34\_43\_inh}=0.04$ and $\beta_{34\_43\_inh}=0.004$, the network exhibits a suppression state in which cell $4$ is suppressed by cell $3$, or vice versa, together with suppression between cells $1$ and $2$, as shown in Fig.\ref{fig:DendronotusAllTrajectories}D. The sequence of activity regimes displayed by cells $3$,$4$ in these trajectories closely parallels that observed in HCOs formed by intrinsic tonic spiking neurons of the SiN model in Figs.\ref{fig:PlantHCOGrid}A and B, which likewise exhibit tonic spiking, chaotic bursting, emergent bursting and suppression. 
    
Figure~\ref{fig:DendronotusSi3AlphaBetaSweepsgExcVsGInhGridgElectric0.001} shows a grid of parametric sweeps at $g_{14\_23\_elec}=0.001$, with the cross excitation $g_{41\_32\_exc}$ varied horizontally and the cross inhibition $g_{14\_23\_inh}$ varied vertically. Each individual sweep shows the network dynamics as the synaptic properties $\alpha_{34\_43\_inh}$ and $\beta_{34\_43\_inh}$ of the tonic-spiking HCO formed by cells $3$,$4$ are varied. The sweeps reveal that the synergistic interactions among the different synaptic pathways give rise to several distinct activity regimes, including tonic spiking (red), chaotic bursting (gray), suppression (green), and emergent network bursting with a varying number of spikes per burst. Representative voltage trajectories for these regimes are shown in Fig.\ref{fig:DendronotusAllTrajectories}. Tonic spiking occurs for weak mutual inhibition (at low values of $\alpha_{34\_43\_inh}$), consistent with the intrinsic tonic-spiking behavior of cells $3$,$4$, whereas suppression occurs for strong mutual inhibition (at high values of $\alpha_{34\_43\_inh}$ or low values of $\beta_{34\_43\_inh}$). At intermediate values of $\alpha_{34\_43\_inh}$ and $\beta_{34\_43\_inh}$, broad regions of emergent bursting appear, together with smaller regions of chaotic bursting.

As the cross inhibition is increased moving vertically upward across the grid, the regions supporting emergent bursting decrease in size, while the number of spikes per burst increases. In contrast, as the cross excitation is increased moving horizontally across the grid, the extent of the emergent-bursting regions remains nearly unchanged, while the number of spikes per burst gradually decreases. The corresponding grid for $g_{14\_23\_elec}=0.002$ is shown in Fig.~\ref{fig:DendronotusSi3AlphaBetaSweepsgExcVsGInhGridgElectric0.002}, and exhibits similar effects of cross excitation and cross inhibition. Increasing $g_{14\_23\_elec}$ from Fig.~\ref{fig:DendronotusSi3AlphaBetaSweepsgExcVsGInhGridgElectric0.001} to Fig.~\ref{fig:DendronotusSi3AlphaBetaSweepsgExcVsGInhGridgElectric0.002} reduces, and in some regions eliminates, the parameter domains supporting emergent bursting, while increasing the number of spikes per burst. Thus, emergent bursting is supported only over restricted ranges of cross inhibition, cross excitation, and electrical coupling, demonstrating that the network rhythm depends on a balance among these interactions rather than simply increasing their individual strengths.

\subsection{The role of excitation and inhibition in emergent CPG bursting}

\begin{figure}[ht!]
\includegraphics[width=.9\textwidth]{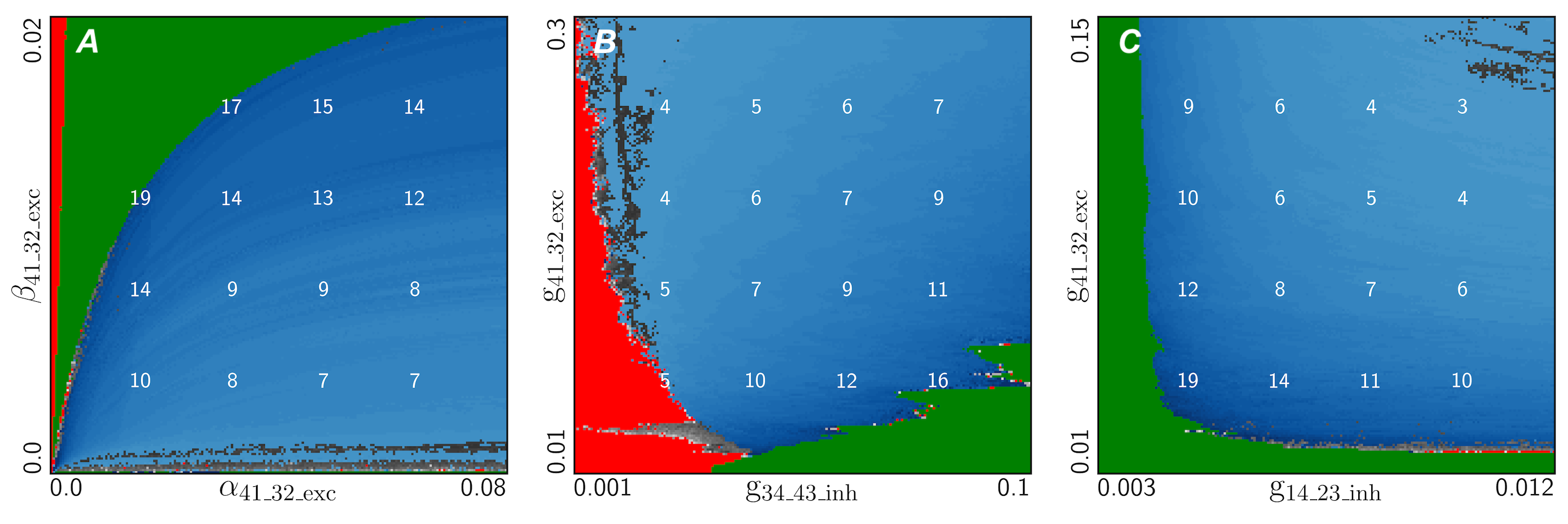}
\caption{ Effects of excitatory and inhibitory synaptic parameters on network dynamics,
with all other model parameters held fixed.
(A) Two-parameter sweep of the slow cross-excitatory synaptic parameters
$\alpha_{41\_32\_{\rm exc}}$ and $\beta_{41\_32\_{\rm exc}}$.
For weaker cross excitation, the network predominantly exhibits tonic
spiking (red) or suppression (green). Emergent bursting appears as the
excitatory activation rate $\alpha_{41\_32\_{\rm exc}}$ increases.
(B) Two-parameter sweep of the mutual-inhibitory conductance between the
tonic-spiking cells, $g_{34\_43\_{\rm inh}}$, and the cross-excitatory
conductance, $g_{41\_32\_{\rm exc}}$. At low values of
$g_{41\_32\_{\rm exc}}$, the network produces only tonic-spiking or
suppressed states. Increasing the cross-excitatory conductance gives rise to
emergent bursting over an intermediate range of mutual-inhibitory strengths.
(C) Two-parameter sweep of the cross-inhibitory conductance
$g_{14\_23\_{\rm inh}}$ and the cross-excitatory conductance
$g_{41\_32\_{\rm exc}}$. Weak cross excitation or cross inhibition favors
suppression, represented by the green region. Emergent bursting occurs at
intermediate values of both conductances. Within the bursting region, the
number of spikes per burst generally decreases as the synaptic conductances
increase. Together, these parameter sweeps demonstrate that emergent
network bursting requires a balance between cross excitation and inhibitory
coupling rather than simply maximal synaptic strength.}\label{fig:DendronotusMiscSweeps}
\end{figure}

We end this section by examining the individual roles of cross excitation, cross inhibition and the mutual inhibition between the tonic-spiking HCO cells 3 and 4, while keeping all other parameters fixed. It was shown in Ref.~\citep{sakurai2016central} that cross excitation plays an important role in emergent bursting and that removing cross excitation can suppress network bursting. Fig.\ref{fig:DendronotusMiscSweeps}A shows a biparametric sweep in which the slow synaptic properties $\alpha_{41\_32\_exc}$ and $\beta_{41\_32\_exc}$ of this cross excitation are varied. For weak excitatory synapses, the network exhibits tonic spiking (red) and suppression (green), whereas emergent bursting appears as the excitatory synapses become stronger through increasing $\alpha$. This role of cross excitation is further illustrated in Fig.\ref{fig:DendronotusMiscSweeps}B, which shows a biparametric sweep of the mutual inhibitory strength between the tonic spikers $g_{34\_43\_inh}$ versus the cross excitation $g_{41\_32\_exc}$. For zero or low values of $g_{41\_32\_exc}$, increasing the mutual inhibition $g_{34\_43\_inh}$ produces only tonic spiking or suppression. This is consistent with the intrinsic tonic-spiking properties of these cells and with the parametric sweep of the tonic-spiker HCO in Fig.~\ref{fig:DendronotusSi3CaShiftvsXShift1_Si3HCO}B. As $g_{41\_32\_exc}$ is increased, emergent bursting appears over an intermediate range of mutual inhibition.

Figure~\ref{fig:DendronotusMiscSweeps}C examines the combined roles of cross inhibition $g_{14\_23\_inh}$ and cross excitation $g_{41\_32\_exc}$. Weak values of either coupling favor suppression (green), whereas emergent bursting occurs over intermediate ranges of both synaptic strengths. Within the bursting region, the number of spikes per burst generally decreases as these conductances increase. Together, these sweeps demonstrate that neither excitation nor inhibition alone is sufficient to account for the emergent rhythm; rather, network bursting requires an appropriate balance between cross excitation and inhibitory coupling.

\section{Conclusions and future directions}

We demonstrated that a four-cell network modeling the {\it Dendronotus iris} swim central pattern generator   can generate robust bursting even when none of its constituent neurons and neither of its isolated HCO-like modules can sustain a burst rhythm. This result goes beyond the familiar observation that mutual inhibition can recruit endogenously non-bursting cells into an HCO: in the \textit{Dendronotus} parametrization, the corresponding pairwise mechanism is insufficient, and bursting emerges only at the next level of circuit assembly. Cross excitation, cross inhibition, and rectified electrical coupling act synergistically to generate this higher-order collective state of the CPG network. The performed parameter sweeps show that the bursting state occupies extended regions of parameter space and reveal systematic changes in the size of these regions and in the number of spikes per burst as the chemical and electrical interactions are varied. Thus, the emergent rhythm is not restricted to a single finely tuned combination of cellular and synaptic parameters, but persists over finite domains in which several network interactions are appropriately balanced.

This hierarchical organization distinguishes several fundamentally different mechanisms of rhythm generation. At the cellular level, an isolated neuron may burst intrinsically as a consequence of its own slow and fast ionic dynamics. At the next level, two neurons that do not burst intrinsically can nevertheless form an oscillatory HCO through reciprocal inhibition. The SiN-model HCO provides such a lower-order example and identifies a slow--fast release mechanism capable of producing conditional pairwise bursting. In contrast, under the \textit{Dendronotus} parametrization, this capability is lost at the HCO level: the isolated tonic-spiking HCO produces only tonic spiking or suppression, while the second HCO is composed of quiescent cells. Bursting is recovered only when these nonbursting modules interact within the complete circuit. The resulting rhythm is therefore not simply inherited from a constituent neuron or from an oscillatory subnetwork, but is generated by the organization of interactions at the level of the assembled network. This provides a concrete example of hierarchical emergence in which a dynamical capability absent from each lower-order subsystem appears at a higher level of circuit organization. This interpretation is consistent with explicit distinction among intrinsic, HCO-emergent, and higher-order network-emergent bursting. 

The parameter sweeps further show that the different synaptic pathways do not merely switch bursting on or off independently. Instead, they reshape the region in which the collective rhythm exists and regulate quantitative properties of that rhythm. Cross excitation and cross inhibition determine whether the interaction between the two HCO-like modules can recruit the network into bursting, while changes in mutual inhibition and rectified electrical coupling alter both the extent of the bursting region and the number of spikes generated during each burst. The disappearance of bursting when these interactions move outside appropriate ranges indicates that the relevant dynamical quantity is not simply the strength of any individual connection, but the balance among multiple chemical and electrical interactions. In this sense, the circuit rhythm is a property of the coupled architecture rather than of any single synaptic pathway. The performed bi-parametric sweeps specifically support this balance interpretation: emergent bursting appears over intermediate ranges of excitatory and inhibitory coupling rather than at maximal coupling strengths. 

Event-based symbolic encoding provides a practical means of resolving this hierarchy across multidimensional parameter spaces. Rather than selecting a few representative trajectories, the symbolic approach allows large parameter domains to be partitioned according to recurring network activity patterns while retaining experimentally meaningful observables such as the number of spikes per burst. This makes it possible to identify candidate regions of tonic spiking, suppression, regular bursting, and more complex activity before applying more computationally intensive dynamical analyses. At the same time, the interpretation of these maps must remain appropriately limited. Equal symbolic words identify the same coarse-grained periodic output under the chosen partition, while LZ shading characterizes the complexity of the resulting symbolic sequence. Neither, by itself, establishes topological equivalence of the underlying attractors, deterministic chaos, dynamical instability, or a particular bifurcation mechanism. The symbolic maps should therefore be viewed as a global exploratory framework for locating and organizing dynamical regimes rather than as a substitute for local bifurcation and stability analysis. This distinction is also explicit in the methodological framework of the paper. 

These results suggest direct experimental tests. Dynamic-clamp perturbations that remove or independently vary cross excitation, cross inhibition, and rectified electrical coupling should collapse, expand, or reshape the bursting regions in the ways predicted by Figs.~\ref{fig:DendronotusSi3AlphaBetaSweepsgExcVsGInhGridgElectric0.001}--\ref{fig:DendronotusMiscSweeps}. Such experiments could test not only whether a particular connection is required for emergent network bursting, but also whether graded changes in its effective strength or kinetics produce the predicted changes in burst organization and spike number. In particular, the model predicts that perturbing one interaction can alter the range over which the remaining interactions support bursting, providing a more stringent test of network synergy than simply deleting individual connections. The experimentally relevant prediction is therefore not only the existence or absence of bursting, but also the geometry and size of the parameter region over which the rhythm can be maintained stably.

Error-function optimization provides a complementary approach for comparing modeled and recorded voltage traces and training blended rhythmic networks \citep{bourahmah2024error}. Parameter optimization and symbolic parameter mapping address different aspects of the same problem: optimization can adequately locate parameter combinations that reproduce selected experimental setups, whereas symbolic sweeps can determine how that solution is embedded within the surrounding dynamical landscape. Combining these approaches could therefore distinguish parameter sets that merely reproduce a target waveform from those that lie inside broader regions supporting the same qualitative network function.

Future work should combine such perturbations with continuation and stability analysis of plausible transitions between representative types of neural activity. In particular, continuation of equilibria and periodic orbits across selected parameter boundaries could determine which transitions correspond to genuine local bifurcations and which reflect changes within a single or among coexisting attractors, if any,  or more complex global mechanisms. Stability calculations would likewise clarify how the tonic-spiking, suppression, and emergent bursting states gain or lose stability as the chemical and electrical interactions are varied. Our analysis reveals the local dynamical structure underlying the network activity maps generated by symbolic parameter sweeps. 

More generally, our findings emphasize one more time that the dynamical function of a neural circuit cannot necessarily be inferred from the dynamics of its isolated components. Neurons that do not burst can participate in and contribute to an HCO that does generate alternating bursts, or form an HCO that cannot produce bursting but can participate in a larger circuit that nevertheless generates robustly emergent  bursting through the network hysteresis mechanism \cite{scully2026slow,scully2024pair} due to the qualitative distinction in functions of post-synaptic neurons whose properties are qualitatively affected by unidirectional synaptic drive flowing in from [temporarily unperturbed] pre-synaptic neurons.  In the given case, the progression from individual cells to HCOs, and eventually to the complete four-cell network provides a particularly transparent example of this principle. Various approaches including Symbolic parameter sweeps, slow--fast analysis, and parameter continuation should be all viewed as complementary tools for resolving different levels of this hierarchy: symbolic methods locate collective dynamical regimes across larger parameter spaces, slow--fast dissection geometry identifies possible candidate bifurcation mechanisms in individual models of interneurons, while bifurcation and stability analyses establish how the dynamical principles can collectively underly resilient rhythm-generation, and manage various transitions between biologically plausible outcomes in small polymorphic neural networks. 

\section*{Conflict of Interest Statement}

The authors declare that the research was conducted in the absence of any commercial or financial relationships that could be construed as a potential conflict of interest.

\section*{Author Contributions}

KP and AS conceptualized the research, developed the methodology, conducted the circuit reconstructions, and wrote the original draft of the manuscript. KP adapted the symbolic encoding software, performed numerical simulations and biparametric sweeps. HJ contributed to the phase space analysis and visualization of the neural manifolds. AS supervised the study, provided laboratory resources, and edited the manuscript. All authors contributed to the article and approved the submitted version.

\section*{Funding}
This research was partially funded by the National Science Foundation award IOS-1455527 and the National Science Foundation award DMS-2407999.  

\section*{Acknowledgments}
We thank the members of the Shilnikov NeurDS Lab for many valuable discussions and suggestions, and Drs.\ P.~S.~Katz and A.~Sakurai for sharing their insights into sea-slug neurophysiology.


\section*{Data Availability Statement}
All the methods and code developed are openly available at Deterministic Chaos Prospector (\url{https://bitbucket.org/pusuluri_krishna/deterministicchaosprospector/})


\section{Appendix: SiN-model: a basic Hodgkin-Huxley type representation of the swim CPG interneurons} \label{sec:appendix1}

The membrane potential, $V$, evolves according to
\begin{equation}
C_{m} {V}^\prime = -I_{I} - I_K - I_{T}  - I_{KCa} -  I_{h}  - I_{leak}  - I_{syn}.  \label{a1}
\end{equation}
The fast inward sodium and calcium $I_{I}$-current is given by\
\begin{equation}
I_{I}=g_{I}\,h\, m^{3}_{\infty}(V)(V-E_{I}),
\end{equation}
where the reversal potential is $E_{I}=30$mV and the maximal conductance is $g_{I}=4$nS. The steady-state activation is described by\
\begin{align}
m_{\infty}(V) &= \frac{\alpha_{m}(V)}{\alpha_{m}(V) +\beta_{m}(V)}, \qquad  
\alpha_{m}(V) = 0.1 \frac{50-V_s}{-1+e^{(50-V_s)/10}}\ , \quad \beta_{m}(V) = 4 e^{(25-V_s)/18},\
\end{align}
while the dynamics of the inactivation gating variable $h$ are given by\
\begin{align}
{h}^\prime =  \frac{h_{\infty}(V)-h}{ {\tau_{h}(V)} }, \quad \mbox{where} \quad 
h_{\infty}(V) = \frac{\alpha_{h}(V)}{\alpha_{h}(V) +\beta_{h}(V)}\quad \mbox{and} 
\quad \tau_{h}(V) = \frac{12.5}{\alpha_{h}(V) +\beta_{h}(V)}
\end{align}
with
\begin{align}
\alpha_{h}(V) = 0.07 e^{(25-V_s)/20}, \quad 
\beta_{h}(V) = \frac{1}{1+e^{(55-V_s)/10}}, \quad \mbox{and} \quad 
V_s = \frac{127V+8265}{105}{\rm mV}.
\end{align}

The fast potassium $I\_K$-current is given by\
\begin{equation}
I\_K=g_{K} n^{4}(V-E\_{K}),\
\end{equation}
where the reversal potential is $E\_{K}=-75$mV and the maximal conductance is $g\_K=0.3$nS. The dynamics of the corresponding gating variable are described by
\begin{align}
{n}^\prime = \frac{n\_{\infty}(V)-n}{\tau\_{n}(V)}, \quad n_{\infty}(V) = \frac{\alpha\_{n}(V)}{\alpha_{n}(V) +\beta_{n}(V)}, \quad  
\tau_{n}(V) = \frac{12.5}{\alpha_{n}(V) +\beta_{n}(V)}, \\ \mbox{where} \quad 
\alpha_{n}(V) = 0.01 \frac{55-V\_s}{e^{(55-V\_s)/10}-1},  \quad
\beta_{n}(V) = 0.125 e^{(45-V\_s)/80}.
\end{align}

The depolarizing h-current is given by
\begin{equation}
I_{h}=g_{h} \frac{y (V-E_{h})}{(1+e^{-(V-63)/7.8)})^3},
\end{equation}
where $E_{h}=70mV$ and $g_h=0.0006nS$. The dynamics of its $y$-activation variable are described by
\begin{equation}
{y}^\prime = \frac{1}{2} \left [ \frac{1}{1+e^{10(V-50)}}-y \right ]/ \left [ 7.1+\frac{10.4}{1+e^{(V+68)/2.2}} \right ],
\end{equation}
whereas the leak current is given by
\begin{align}
I_{L} &=g_{L} (V-E_{L}),
\end{align}
with $E_{L} =-40{\rm mV}$ and $g_{L}=0.003{\rm nS}$.

The slow-current subsystem includes the TTX-resistant sodium and calcium $I_{T}$-current, given by
\begin{align}
I_{T} &= g_{T} x (V-E_{I}),
\end{align}
with $E_{I} =30mV$ and $g_{T}=0.01nS$. The dynamics of its slow activation variable are described by\
\begin{align}
{x}^\prime = \frac{x_{\infty}(V)-x}{\tau_{x}}, \quad \mbox{where} 
\quad x_{\infty} = \frac{1}{1+e^{-0.15(V+50-{\rm x_{shift}})}},
\end{align}\label{eqttx}
with the time constant $\tau_{x}$ set to 100 in this study.

The slowest outward $Ca^{2+}$ activated $K^+$ current is given by\
\begin{equation}
I_{KCa} =g_{KCa}\frac{[Ca]_i}{0.5+[Ca]_i}(V-E_{K})
\label{IKCa}
\end{equation}
with $E_{K}=-75\mbox{mV}$ and $g_{KCa}=0.03$. The dynamics of the intracellular calcium concentration are governed by
\begin{equation}
{[Ca]}^\prime  =  \rho \left ( K_{c}\, x\, (E_{Ca}-V + {\rm Ca_{shift}})-[Ca] \right ),  \label{eqCa}
\end{equation}
where the Nernst reversal potential is $E\_{Ca}=140$mV, and the small constants are $\rho= 0.0003{\rm ms}^{-1}$ and $K_c=0.0085{\rm mV}^{-1}$.


\end{document}